\documentclass{article}

\usepackage{microtype}
\usepackage{graphicx}
\usepackage{bbding}
\usepackage{subcaption}
\usepackage{booktabs} 
\usepackage{makecell}
\usepackage{hyperref}
\usepackage{afterpage}   
\usepackage{placeins}
\usepackage{ulem}
\usepackage[most]{tcolorbox}

\usepackage[nolinenum]{icml2026}

\usepackage{amsmath}
\usepackage{amssymb}
\usepackage{mathtools}
\usepackage{amsthm}
\usepackage{enumitem}
\usepackage{multirow}
\usepackage{bm}
\usepackage{colortbl}
\usepackage{array}
\usepackage{color}
\definecolor{boxheader}{RGB}{50, 60, 75}   
\definecolor{boxbg}{RGB}{250, 250, 250}    
\definecolor{accent}{RGB}{0, 90, 150}      
\definecolor{jsonkey}{RGB}{156, 39, 176}   
\definecolor{jsonstring}{RGB}{56, 142, 60} 
\definecolor{jsonnum}{RGB}{21, 101, 192}   

\newtcolorbox{promptbox}[1]{
  enhanced,
  breakable,
  colback=boxbg,
  colframe=boxheader,
  colbacktitle=boxheader,
  coltitle=white,
  fonttitle=\bfseries\Large,
  title={#1},
  arc=3mm,           
  boxrule=1.2pt,     
  drop shadow=black!30!white, 
  left=6mm, right=6mm, top=4mm, bottom=4mm, 
  toptitle=2mm, bottomtitle=2mm, 
}

\usepackage[capitalize,noabbrev]{cleveref}

\theoremstyle{plain}

\theoremstyle{definition}

\theoremstyle{remark}

\usepackage[textsize=tiny]{todonotes}

\icmltitlerunning{SRBench}

\begin{document}

\twocolumn[
  \icmltitle{SRBench: A Comprehensive Benchmark for \\ Sequential Recommendation with Large Language Models}



  \icmlsetsymbol{equal}{*}

  \begin{icmlauthorlist}
    \icmlauthor{Jianhong Li}{yyy}
    \icmlauthor{Zeheng Qian}{yyy}
    \icmlauthor{Wangze Ni}{yyy}
    \icmlauthor{Haoyang Li}{yyy}
    \icmlauthor{Hongwei Yao}{yyy}
    \icmlauthor{Yang Bai}{yyy}
    \icmlauthor{Kui Ren}{yyy}
  \end{icmlauthorlist}

  \icmlaffiliation{yyy}{Zhejiang University, Hangzhou, China}

  \icmlcorrespondingauthor{Wangze Ni}{niwangze@zju.edu.cn}

  \icmlkeywords{Machine Learning, ICML}

  \vskip 0.3in
]



\printAffiliationsAndNotice{}  
\begin{abstract}
LLM development has aroused great interest in Sequential Recommendation (SR) applications.  However, comprehensive evaluation of SR models remains lacking due to the limitations of the existing benchmarks: 1) an overemphasis on accuracy, ignoring other real-world demands (e.g., fairness); 2) existing datasets fail to unleash LLMs' potential, leading to unfair comparison between Neural-Network-based SR (NN-SR) models and LLM-based SR (LLM-SR) models; and 3) no reliable mechanism for extracting task-specific answers from unstructured LLM outputs. To address these limitations, we propose SRBench, a comprehensive SR benchmark with three core designs: 1) a multi-dimensional framework covering accuracy, fairness, stability and efficiency, aligned with practical demands; 2) a unified input paradigm via prompt engineering to boost LLM-SR performance and enable fair comparisons between models; 3) a novel prompt-extractor-coupled extraction mechanism, which captures answers from LLM outputs through prompt-enforced output formatting and a numeric-oriented extractor. We have used SRBench to evaluate 13 mainstream models and discovered some meaningful insights (e.g., LLM-SR models overfocus on item popularity but lack deep understanding of item quality). Concisely, SRBench enables fair and comprehensive assessments for SR models, underpinning future research and practical application.
\end{abstract}

\section{Introduction}
\label{sec:Intro}
Inspired by the excellent contextual understanding capabilities of LLMs, SR performance via LLMs has drawn significant attention from academia (e.g., ICML community \citep{huangimproving,forouzandehsharp}) and industry (e.g., Amazon \citep{wang2020time}). However, researchers also have expressed concerns about the practical issues of LLM-SR models, like stability and fairness \citep{zhang2023chatgpt,ma2023large}. Thus, it is necessary to design a comprehensive benchmark that systematically evaluates and compares the performance of NN-SR and LLM-SR models.   

Existing benchmarks have evaluated models in partial models or evaluation dimensions. LLMRec~\cite{liu2023llmrec} evaluates only LLM-SR models on recommendation tasks, without comparing NN-SR models. OpenP5~\cite{xu2024openp5} and RecBench~\cite{liu2025benchmarking} cover NN-SR and LLM-SR models yet only use accuracy-based evaluation metrics. Beyond accuracy, real-world SR scenarios involve other crucial dimensions (e.g., fairness)~\cite{Rampisela2023EvaluationMO}, which are also vital for judging SR quality and rationality. Overall, existing benchmarks are insufficient for a unified and holistic evaluation.

\begin{figure*}[!t]
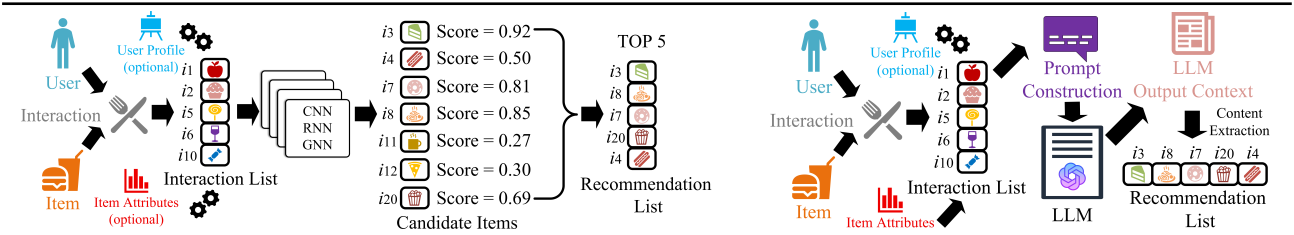

\vspace{-0.65em}     
    \centering
    \begin{subfigure}[c]{0.55\textwidth}
        \centering
        \includegraphics[width=\textwidth]{process1.png}
        \caption{
        The NN-SR model learns user's dynamic preference from the historical interaction sequence, and predicts the candidate items.}
        \label{fig:nn-sr}
    \end{subfigure}
    \hspace{1em}
    \begin{subfigure}[c]{0.4\textwidth}
        \centering
        \includegraphics[width=\textwidth]{process2.png}
        \caption{
        LLM-SR model predicts SR results based on constructed prompt.
}        
        \label{fig:llm-sr}
    \end{subfigure}
        \hfill 
    \caption{Frameworks for Sequential Recommendation: 
     (a) NN-SR methods and (b) LLM-SR methods.}
    \label{fig1}
\vspace{-1em}
\end{figure*}  

Thus, we need to solve a technical question: how to design a benchmark to comprehensively evaluate and compare both NN-SR and LLM-SR models? However, the question is difficult, and three challenges need to be solved:

\begin{itemize} [leftmargin=1em]
\vspace{-0.6em}
\item \textbf{Challenge 1: Multi-Dimensional Evaluation.} NN-SR models capture sequential patterns in user profiles and historical interactions, while LLM-SR models use pretrained knowledge for semantic reasoning. Their distinct foundations, along with varied practical needs in sequential recommendation, make it challenging to define evaluation dimensions and metrics that apply to both model paradigms yet effectively differentiate their performance.

\item \textbf{Challenge 2: Input Paradigm Alignment.} As shown in Figure ~\ref{fig1}, NN-SR and LLM-SR employ distinct input paradigms, with LLM-SR requiring additional natural language prompts. Moreover, prompt engineering significantly affects LLM-SR performance~\cite{xu2024openp5}. Thus, designing an effective unified input paradigm to convert structured datasets into LLM-SR-compatible prompts and enhance such prompts to unleash LLMs’ potential for fair comparison is challenging.
     
\item \textbf{Challenge 3: Output Standardization Mechanism.} LLM-SR models output results in natural-language form, which is unstructured and disordered. Moreover, due to LLMs' inherent randomness, LLMs' outputs may contain hallucinations, which makes it harder to accurately extract answers for validation.
\end{itemize}

\vspace{-0.7em}
\indent In this paper, we propose SRBench to address the aforementioned challenges and make the contributions:
\begin{itemize} [leftmargin=1em]
\vspace{-0.2em}
\item \uline{To overcome Challenge 1}, we conduct a comprehensive in-depth survey to analyze real-world SR demands. Based on NN-SR and LLM-SR model features, we propose a multi-dimensional benchmark framework with four important dimensions: accuracy, fairness, stability and efficiency. For each dimension, output-only metrics are designed to ensure compatibility with all models and enable holistic, unbiased SR evaluation. (Sec. \ref{multi-dimensionalmetrics})

\item \uline{To solve Challenge 2}, we propose a unified prompt-enhanced input paradigm that converts structured sequential data for NN-SR models into prompts for LLM-SR models. We extract historical interaction items from the sequential data and populate them chronologically into a tailored prompt template, thereby ensuring consistent evaluation data input for models. This prompt is further enhanced through role-playing instructions and detailed semantic interpretations of the extracted data to fully leverage the reasoning capabilities of LLMs. (Sec.~\ref{dataconversion})

\item \uline{To address Challenge 3}, we propose a prompt-extractor-coupled mechanism to extract SR results from unstructured LLM-SR responses. Input prompts are enhanced through prompt engineering to produce semi-structured outputs. The extractor module identifies numerical values in LLM-SR outputs as item IDs, and with such IDs as anchors, further converts LLM-SR outputs into sequential formats fully aligned with NN-SR outputs following the identified ID sequence. (Sec.~\ref{content extraction}) 

\item \uline{For practical assessment}, we evaluate 13 SR models of both types in multiple datasets, revealing multiple key insights such as the popularity-driven exposure tendency and the stability defect of LLM-SR models. (Sec.~\ref{experiment})
\end{itemize}

In addition, we discuss the related works in Sec.~\ref{relatedwork} and conclude the paper in Sec.~\ref{conclusion}.

\section{Related Work}
\label{relatedwork}   

\label{sec 2.1}
\uline{NN-SR Models}. NN-SR models rely on explicit sequential modeling over structured user-item interaction data and can be further grouped into three representative subcategories: \textbf{1) CNN-based methods}, such as Caser \cite{tang2018personalized} and its extensions \cite{yan2019cosrec,chen2022double,yuan2019simple}, adapt convolutional kernel operators to extract local or short-range sequential patterns from fixed interaction windows, enabling modeling of users’ short-term interests and improving SR performance. \textbf{2) RNN-based approaches}, including GRU-based models \cite{hidasi2018recurrent,donkers2017sequential} are designed to capture long-term dependencies in interaction sequences, with the goal of learning expressive temporal user preference representations. \textbf{3) Graph-based methods}, which further extend sequential modeling by explicitly incorporating higher-order user-item relationships. Representative works such as NGCF \cite{wang2019neural} and LightGCN \cite{he2020lightgcn,lee2024revisiting} propagate collaborative signals over interaction graphs, while recent works explore dynamic graph structures to improve model temporal evolution and user behavior change \cite{chang2021sequential,zhang2022dynamic,zheng2022ddghm,zhang2023learning,yu2023towards}.

\uline{LLM-SR models}. For LLM-SR models, LLMs are explored as the backbone for SR, leading to a paradigm shift from explicit sequence modeling to prompt-based semantic reasoning. Under this paradigm, prompt-based methods, such as Chat-REC \cite{gao2023chat}, NIR \cite{wang2023zero}, and related approaches \cite{geng2022recommendation,lyu2023llm,xi2024towards,silva2024leveraging,liu2023chatgpt}, reformulate user interaction histories into natural-language prompts to guide LLMs in generating SR results. Other studies further adopt parameter-efficient fine-tuning techniques, such as LoRA \cite{hu2021lora}, to adapt LLMs for recommendation tasks \cite{bao2023tallrec,yue2023llamarec}. 

\uline{Benchmarks for SR}. Owing to the fundamental differences between NN-SR and LLM-SR models, several benchmarks have been proposed to evaluate SR under emerging LLM-based approaches. OpenP5 \cite{xu2024openp5} provides a unified platform for benchmarking generative SR, with an emphasis on prompt construction and item indexing. LLMRec \cite{liu2023llmrec} evaluates LLM performance across multiple recommendation tasks by transforming structured data into natural-language prompts. Jiang et al. ~\cite{jiang2025beyond} further investigate LLMs as ranking models in instruction-driven evaluation mode. Other benchmarks, such as RecBench \cite{liu2025benchmarking}, explore the impact of different item representation strategies, including IDs, text, and embeddings. In addition, FairEvalLLM \cite{deldjoo2024normative} and IFairLRS \cite{deldjoo2024understanding} extend evaluation from accuracy to fairness-related considerations.

However, as Table \ref{tab:comp dimension} presents, these benchmarks overemphasize accuracy metrics, while neglecting other metrics in real-world deployment. Specifically, dimensions of accuracy, fairness, stability and efficiency respectively characterize user-level effectiveness, system-level behavior, behavioral reliability, and deployment feasibility, forming a complete and non-overlapping evaluation space for real-world SR performance. Additionally, although existing datasets and metrics design are fit for NN-SR models, they can't be directly applied to LLMs, as Sec. \ref{sec 2.1} discussed. Thus, current SR models require a comprehensive benchmark that not only preserves the key evaluation dimensions but also aligns input paradigms and standardizes output extraction to enable fair and meaningful comparisons across model paradigms.

\begin{table}[t]
\centering
\caption{Comparison of Dimensions in Different Benchmarks.}
\label{tab:comp dimension}
\resizebox{\columnwidth}{!}{
\begin{tabular}{lcccc}
\toprule
\textbf{Input Methods} & \textbf{Accuracy} & \textbf{Fairness} & \textbf{Stability} & \textbf{Efficiency} \\
\midrule
OpenP5 & \color{green}{\Checkmark} & \color{red}{\XSolidBrush} & \color{red}{\XSolidBrush} & \color{red}{\XSolidBrush} \\
LLMRec & \color{green}{\Checkmark} & \color{red}{\XSolidBrush} & \color{red}{\XSolidBrush} & \color{red}{\XSolidBrush} \\
RecBench & \color{green}{\Checkmark} & \color{red}{\XSolidBrush} & \color{red}{\XSolidBrush} & \color{red}{\XSolidBrush} \\
FairEvalLLM & \color{red}{\XSolidBrush} & \color{green}{\Checkmark} & \color{red}{\XSolidBrush} & \color{red}{\XSolidBrush} \\
IFairLRS & \color{green}{\Checkmark} & \color{green}{\Checkmark} & \color{red}{\XSolidBrush} & \color{red}{\XSolidBrush} \\
\midrule
SRBench & \color{green}{\Checkmark} & \color{green}{\Checkmark} & \color{green}{\Checkmark} & \color{green}{\Checkmark} \\
\bottomrule
\end{tabular}
}
\vspace{-2em}
\end{table}

\section{SRBench}
\label{srbench}

To address the challenges summarized in Section \ref{sec:Intro}, we propose SRBench, a unified benchmark to tackle these challenges via three core components: multi-dimensional metrics, unified input paradigm, and prompt-extractor-coupled extraction mechanism.

\subsection{Multi-dimensional Metrics}
\label{multi-dimensionalmetrics}

In SR scenarios, the performance of SR models can't be characterized by accuracy metrics \cite{Jadon2023ACS} alone. Practical deployment spins off diverse requirements on SR models, including result relevance, exposure fairness, behavior consistency, and computational efficiency. 

To this end, we propose a multi-dimensional evaluation that decomposes SR performance into several complementary dimensions, each capturing a distinct aspect of model performance. Rather than relying on a single metric, we operate each dimension through concrete and measurable metrics that reflect observable behaviors of SR models in practice. Specifically, our benchmark evaluates SR performance along with four key dimensions: \textit{Accuracy}, \textit{Fairness}, \textit{Stability}, and \textit{Efficiency}. Within this framework, we propose new metrics in several dimensions to enable a more comprehensive assessment across different SR paradigms.

\textbf{Accuracy.} Recommendation accuracy measures the model’s ability to correctly recommend and rank relevant items for users. We adopt standard top-$k$ ranking metrics, including \textit{Recall@K} \cite{Sajjadi2018AssessingGM} and \textit{NDCG@K} \cite{Jrvelin2000IREM}, to evaluate whether the ground-truth items appear in the recommended list and how well they are ranked. These metrics are widely used in both NN-SR and LLM-SR models, enabling fair comparison. While accuracy captures whether relevant items are correctly recommended and ranked, it can't depict the overall effectiveness of SR models in practical deployment. Thus, SRBench treats accuracy as a necessary but insufficient dimension, and complements it with additional metrics.

\textbf{Fairness.} Fairness evaluates whether SR models provide equitable exposure across items and users. In real-life LLM-SR scenarios, bias like undue exposure preference or quality defect may lead to negative outcomes like price discrimination, over-advertising, even hard sell. To address this, based on fairness research \cite{Rampisela2023EvaluationMO,Leonhardt2018UserFI}, SRBench divides fairness into complementary perspectives that reflect both \uline{item-side exposure and user-side recommendation quality}. We present two metrics to depict how SR models distribute exposure and whether users receive high-quality SR results.   

\uline{From item perspective}, SRBench adopts the widely used \textit{Average Recommendation Popularity (ARP)} \cite{Bai2025EnsuringAA} to measure exposure bias toward popular items. \textit{ARP} quantifies whether an SR model disproportionately favors items with high historical interaction frequencies and is defined as: $\bm{ARP@K = \frac{1}{U} \sum_{n=1}^{U} \left( \frac{1}{K} \sum_{i=1}^{K} P_i \right)}$, 
where $U$ denotes the number of users, $K$ is the recommendation list length, and $P_i$ represents the historical popularity (total interaction count) of the $i$-th recommended item. Higher \textit{ARP} indicates a stronger preference to popular items, reducing the item fairness. However, \textit{ARP} alone is insufficient to provide a holistic fairness assessment, as it can't measure the quality of recommended items. 

To address this limitation, based on existing studies on quality-rating relation \cite{Guan2013RecommendationAB}, SRBench proposes a \uline{novel user-perspective fairness metric}, \textit{Average Recommendation Quality (ARQ)}, which evaluates fairness by measuring the average quality of delivered items, and it is defined by: $\bm{ARQ@K = \frac{1}{U} \sum_{n=1}^{U} \left( \frac{1}{K} \sum_{i=1}^{K} Q_i \right)}$, 
where $Q_i$ denotes the quality score (e.g., rating) of the $i$-th recommended item. \textit{ARQ} captures whether users receive high-quality SR results, complementing \textit{ARP} by distinguishing item popularity and quality bias. Higher \textit{ARQ} shows that the user is recommended with higher-quality items, enhancing fairness from user's perspective. By jointly considering \textit{ARP} and \textit{ARQ}, SRBench enables a comprehensive fairness evaluation that differentiates between popularity-driven exposure and high-quality, equitable SR behavior.

\textbf{Stability.} While accuracy evaluates relevance of recommended items, and fairness assesses whether exposure and quality are balanced, neither dimension captures if SR behaviors remain consistent. LLM-SR models often exhibit inconsistency due to their inherent randomness. Such unpredictability leads to low model reliability, even when accuracy and fairness metrics appear satisfactory, making stability hard to quantify. To evaluate stability from SR results, based on research of stability \cite{adomavicius2012stability} and variable-control theory, SRBench decomposes stability into complementary perspectives that reflect both \uline{quality-level consistency and output-level consistency}. Specifically, quality-level consistency refers to whether an SR model delivers recommendations in similar quality levels. If the recommended items contains huge quality gap, the model is obviously unstable. Besides, output-level consistency directly measures whether the model outputs similar results across repeated runs with same settings.

\uline{From the quality perspective}, SRBench innovatively presents a stability metric, \textit{Average Recommendation Quality Variance (ARQV)}, which measures the variance of item quality in recommendation lists. \textit{ARQV} evaluates whether an SR model consistently delivers recommendations with similar quality levels, rather than unstable between high-quality and low-quality items due to random generation effects. The formulation of \textit{ARQV} is calculated by: $\bm{ARQV@K = \frac{1}{U} \sum_{n=1}^{U} \mathrm{Var}_{i=1}^{K}(Q_i)}$, 
where $U$ denotes the total number of users, $K$ is the recommendation list length, and $Q_i$ represents the quality score (e.g., rating) of the $i$-th recommended item. $\mathrm{Var}(\cdot)$ computes the variance of item quality in each recommendation list, measuring the stability of recommendation quality.

\uline{From the output perspective}, SRBench proposes \textit{Average Recommendation Repetition (ARR)}, which quantifies SR result consistency by measuring the overlap of recommended items across repeated executions. \textit{ARR} captures whether an SR model repeatedly recommends similar items, revealing output-level reproducibility in SR behavior. The formulation of \textit{ARR} is defined as: $\bm{ARR@K = \frac{1}{T \times K} \sum_{n=1}^{T} Rep_n}$, 
where $T$ denotes the number of repeated SR executions under identical settings, $K$ is recommendation list length, and $Rep_n$ represents the number of overlapping items between SR results in the $n$-th execution and those from other runs. By combining \textit{ARQV} and \textit{ARR}, SRBench measures stability from both quality and output perspectives. A lower \textit{ARQV} indicates more stable recommendation quality, while a higher \textit{ARR} reflects stronger consistency in SR results.


\textbf{Efficiency.} Beyond accuracy, fairness, and stability, computational efficiency is a critical dimension for SR evaluation. In practical deployment scenarios, slow speed of SR process will severely reduce user satisfaction, which forms the need of efficiency evaluation to SR models. To effectively assess efficiency, SRBench focuses on a core metric that quantifies the time cost of SR process. Here, SRBench adopts \textit{Average Recommendation Time (ART)}, which commonly measures the average time for SR models to produce a recommendation list and is defined as follows: $\bm{ART = \frac{1}{N} \sum_{n=1}^{N} T_{c_n}}$, 
where $N$ denotes the total number of SR executions, and $T_{c_n}$ represents the time cost of the $n$-th execution. \textit{ART} captures how efficient an SR model is, reflecting both algorithmic speed and system responsiveness. A lower \textit{ART} value indicates higher efficiency, demonstrating that the model can deliver timely recommendations even under practical constraints. By evaluating \textit{ART} with other dimensions, SRBench provides a holistic view of SR performance, balancing quality, reasonability, reliability, and deployability.

\subsection{Unified Input Paradigm}
\label{dataconversion}

As discussed in \textbf{Challenge 2}, the input paradigm mismatch makes indexed interaction data not only uninterpretable to LLMs, but also leads to unreliable reasoning. To enable fair comparison across model paradigms, we propose a unified input paradigm based on prompt-engineering enhancement. The input paradigm converts structured interaction sequence into standardized natural-language prompts, excluding the interference form prompt-design variations.

\textbf{Datasets.} As foundation, we adopt three representative datasets: \textit{MovieLens-100K}, \textit{Beauty} and \textit{Yelp}. These datasets are commonly used to assess SR performance, covering different scales and domains, whose details and preprocessing are listed in Appendix \ref{dataset}. 

\textbf{Data Conversion.} The backbone of the unified input paradigm is \textit{data conversion}. After analyzing previous LLM-SR prompt formation \cite{xu2024openp5}, we firstly design a general prompt template that contains SR motivation. Then, we extract historical interaction items from user interaction sequence and chronologically embed them to the prompt, which maintains the input consistency between NN-SR and LLM-SR models. However, the prompt only triggers LLM-SR models for basic SR process, which is not enough for exhibiting their best performance and requires enhancement.  

\textbf{Prompt Enhancement.} Based on the data conversion, we achieve \textit{prompt enhancement} by attaching extra information required for LLM-SR process. Specifically, according to the research on LLM reasoning \cite{Ke2025ASO} and prompting strategy \cite{Lu2024LLMDE}, we extend SR objective with role-playing instructions, which proactively activates LLM's specialized knowledge storage in fields of SR. As per the item IDs in user interaction sequence, we additionally attach related detailed semantic item interpretations to the prompt, such as item titles, categories and descriptions, forming one-to-one correspondence, for LLM-SR models to better perform their outstanding semantic reasoning capabilities.

\begin{figure}[t]
    \centering
    \includegraphics[width=0.4\textwidth]{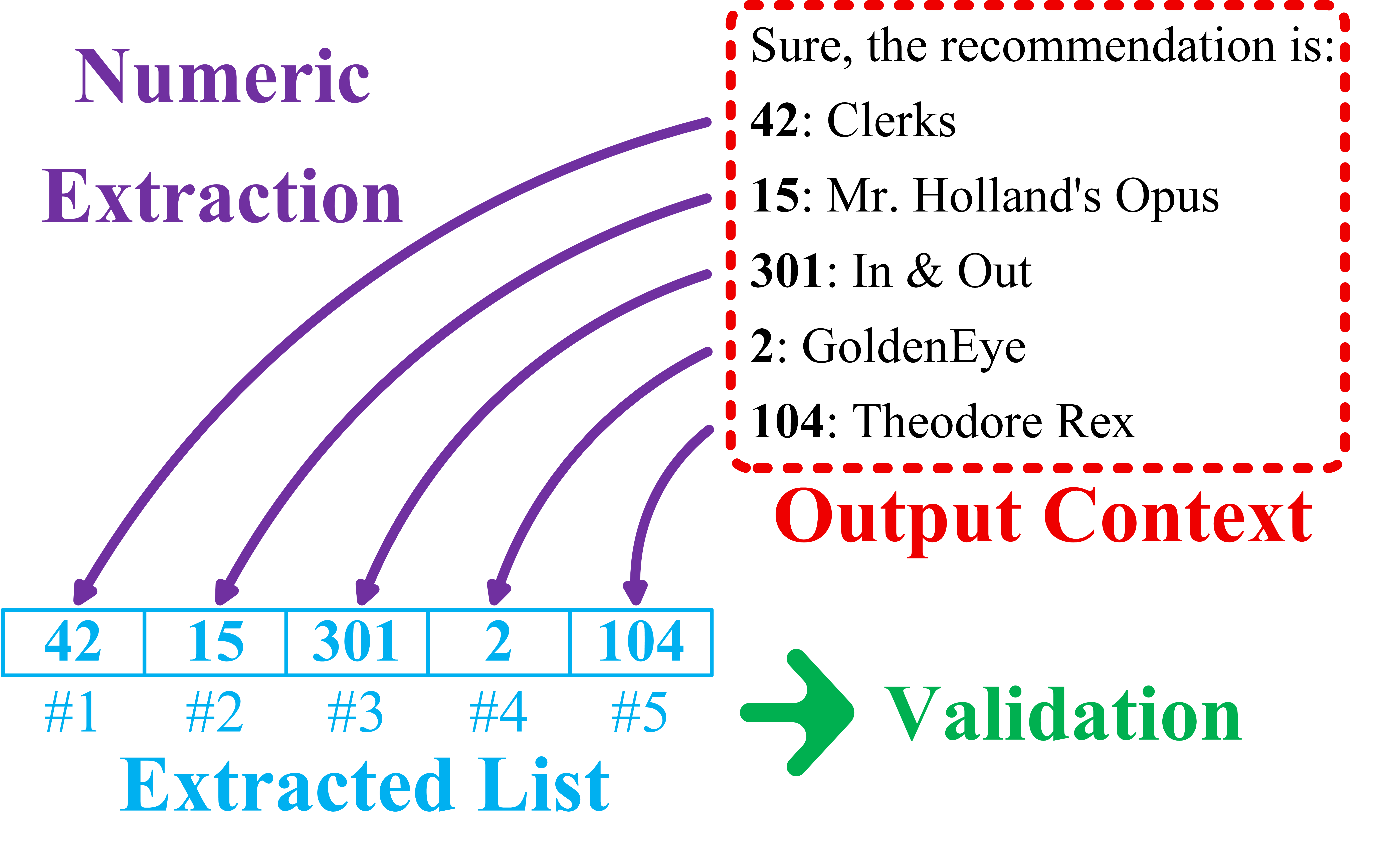}
    \caption{Process of Content Extraction in SRBench.} 
    \label{fig:output}
\vspace{-1.5em}
\end{figure}

\subsection{Prompt-extractor-coupled Extraction Mechanism}
\label{content extraction}

As described in \textbf{Challenge 3}, while NN-SR models recommend structured results that are direct for evaluation, LLM-SR models generate recommendations in free-form natural language even containing hallucination as Table \ref{tab:hallucination_rates} in Appendix \ref{hallu} presents, which can't be directly used for standard evaluation even under explicitly constrained prompts. The missing of a standardized extraction mechanism leads to unreliable metric calculation, failing to reflect the effectiveness of SR models. 

To solve this, based on prompt-enforced output formatting and numeric-oriented extraction method, we implement a prompt-extractor-coupled extraction mechanism to accurately extract SR results from LLM-SR model's natural-language outputs. According to the semi-structured outputs that model generates after taking in enhanced prompt, as shown in Figure \ref{fig:output}, the mechanism automatically identifies numerical values among output context as item IDs. Utilizing these item IDs as anchors, the mechanism transforms LLM-SR output context into sequential format as extracted list, which is tightly aligned with NN-SR model's predicted item ID sequence.

After extraction, the mechanism individually checks every extracted item ID by comparing with item ID's universal set to verify the validity of recommended items. If the item ID doesn't exist in the universal set, the mechanism will confirm it as hallucinated item and remove it from results, ensuring the true reflection of SR performance instead of artifacts from text generation.


\section{Experiment}
\label{experiment}
In this section, we conduct extensive experiments on real-world datasets to answer the following research questions:
 
\textbf{Q1.} Can SRBench be applied to different types of SR models across various datasets, demonstrating universality and accurate evaluation?

\textbf{Q2.} How do LLM-SR models perform compared to NN-SR models across multiple dimensions, such as accuracy, fairness, stability, and efficiency in SRBench?

\textbf{Q3.} Can SRBench's unified input paradigm effectively convert the structured data into a clear and well-guided prompt, enabling LLM-SR models to perform better?

\textbf{Q4.} Does prompt-extractor-coupled extraction mechanism in SRBench robustly extract the recommendation information in LLM output for subsequent validation? 

\textbf{Q5.} How does the length of the interaction sequence influence the performance of LLM-SR models in SRBench?

The evaluated SR models and settings are introduced in Sec.~\ref{es-1}, followed by experimental performance analysis to answer research questions in Sec.~\ref{es-2}. Finally, key insights are summarized in Sec.~\ref{es-3}. 

\subsection{Evaluated SR Models and Settings} 
\label{es-1}
We select and evaluate multiple typical SR models, covering all model paradigms and classes in Sec. \ref{relatedwork}. The NN-SR models are \textit{NCF}, \textit{Caser}, \textit{GRURec}, \textit{NGCF} and \textit{LightGCN}. The LLM-SR models are \textit{Llama-2}, \textit{Llama-3.2}, \textit{GPT-4.1}, \textit{Grok-3}, \textit{DeepSeek-v3}, \textit{GLM-3-turbo}, \textit{Claude-Sonnet-4} and \textit{Claude-Sonnet-4-thinking}. The details of the evaluated SR models are listed in Appendix \ref{evaluated}.

\begin{table*}[t]
\centering
\caption{Multi-dimensional Performance Comparison On Yelp Dataset.}
\label{tab:yelp}
\setlength{\tabcolsep}{4pt}
\begin{tabular}{cccccccc}
\toprule
\multirow{2}{*}{\textbf{Model}} &
\multicolumn{2}{c}{\textbf{Accuracy}} &
\multicolumn{2}{c}{\textbf{Fairness}} &
\multicolumn{2}{c}{\textbf{Stability}} &
\multicolumn{1}{c}{\textbf{Efficiency}} \\
\cmidrule(lr){2-3}\cmidrule(lr){4-5}\cmidrule(lr){6-7}\cmidrule(lr){8-8}
 & Recall@5 & NDCG@5 & ARP & ARQ & ARQV & ARR & ART(s) \\
\midrule
NCF & 0.0031 & 0.0019 & \textbf{11.2577} & \textbf{4.1319} & 0.3295 &  0.7582 & 0.5841 \\
Caser &  0.0153 & 0.0096 & 89.7224 & 4.0276 & \textbf{0.1632} & 0.7010 & 0.5367 \\
GRURec & 0.0295 & 0.0189 & 44.7208 & 3.9194 & 0.3081 & 0.4428 & 0.3701 \\
NGCF & 0.0130 & 0.0080 & 24.2122&3.9608&0.2567&  0.5394 & 0.2563\\
LightGCN & 0.0157 & 0.0097 & 31.0792 & 4.0191 &  0.1978 & \textbf{0.8152} & \textbf{0.2527} \\
\midrule
Llama-2 &  0.0107 &  0.0056 & 80.8269 & 3.7182 & 0.6965 & 0.4500 & 0.5305 \\
Llama-3.2 & 0.0087 & 0.0048 & 70.0576 & 3.5211 & 0.8176 & 0.4500 & 0.2476 \\
GPT-4.1 & 0.0296 & 0.0207 & 157.8304 & 3.8119 & 0.2742 & 0.4129 & 1.3991 \\
Grok-3 & \textbf{0.0656} & \textbf{0.0511} & 168.3210 & 3.7143 & 0.2637 & 0.2810 & 20.6676 \\
DeepSeek-v3 & 0.0332 & 0.0246 & 200.4863 & 3.8200 & 0.2291 & 0.4439 & 1.7944 \\
GLM-3-turbo & 0.0215 & 0.0146 & 153.9287 & 3.7687 & 0.3646 & 0.2821 & 1.7239 \\
Claude-Sonnet-4 & 0.0159 & 0.0128 & 60.6568 & 3.0308 & 0.4235 & 0.4309 & 2.5346 \\
Claude-Sonnet-4-thinking & 0.0076 & 0.0063 & 103.2343 & 3.7165 & 0.3634 & 0.2011 & 9.9303 \\
\midrule
Llama-2 (few-shot-5) & 0.0116 & 0.0057 & 66.7336 &  3.7448 &  0.7316  &  0.4500 & 0.4821 \\
Llama-3.2 (few-shot-5) & 0.0096 & 0.0052 &  53.4442 & 3.5331 & 0.8492 & 0.4500 & 0.2227 \\
GPT-4.1 (few-shot-5) & 0.0368 & 0.0253 & 97.1358 & 3.7907 & 0.4009 & 0.4249 & 1.5025 \\
Grok-3 (few-shot-5) & 0.0820 & 0.0648 & 101.2338 & 3.6908 & 0.4203 & 0.3119 & 15.8779 \\
DeepSeek-v3 (few-shot-5) & 0.0420 & 0.0288 & 150.7522 & 3.8181 & 0.3915 & 0.4457 & 1.4499 \\
GLM-3-turbo (few-shot-5) & 0.0273 & 0.0170 & 105.6973 & 3.7779 & 0.4583 & 0.3041 & 1.4875 \\
Claude-Sonnet-4 (few-shot-5) & 0.0112 & 0.0086 & 46.9631 & 3.6750 & 0.6022 & 0.4342 & 2.8064 \\
Claude-Sonnet-4-thinking (few-shot-5) & 0.0071 & 0.0055 & 84.4376 & 3.5915 & 0.4538 & 0.2270 & 8.9922 \\
\bottomrule
\end{tabular}  
\vspace{1em}
\end{table*}

For specific implementation, all experiments are conducted in the following environment: Operating System is Ubuntu 20.04 and programming language is Python 3.10. Libraries and frameworks are Transformers 4.45.2 and PyTorch 2.3.0. For NN-SR models, we use the $Adam$ optimizer with a learning rate of $e^{-3}$, which is a commonly adopted default configuration for neural network learning. For LLM-SR models, to mostly reduce randomness, the temperature value is set to 0. All experiments were repeatedly run 10 times and the average values were taken.

\subsection{\textbf{Experiment Performance and Analysis}}
\label{es-2}
The benchmark results on the Yelp, ML-100K and Beauty datasets are respectively presented in Table \ref{tab:yelp}, and Table \ref{tab:100k}, Table \ref{tab:beauty} in Appendix \ref{additional}. Due to Yelp dataset's largest data scale, we take results of Yelp dataset for further analysis in Q2, Q3, Q4 and Q5.

\textbf{SRBench across Datasets and Models.} To answer \textbf{Q1}, based on results in Table \ref{tab:yelp}, Table \ref{tab:100k} and Table \ref{tab:beauty}, it is clear that SRBench is capable of evaluating different types of models, including NN-based, local LLM-based and online LLM-based. \uline{SRBench shows universality on various datasets with different structures and indexing orders}. We also compare results from SRBench with previous work. As shown in Figure \ref{tab:100k}, on same MovieLens dataset, Caser's Recall@5 tested by SRBench is 0.0637, which is very close to Caser's own proposed evaluation result (0.0632) \cite{tang2018personalized}. Above consistency demonstrates that \uline{SRBench is capable of accurately evaluating SR models}.  

\begin{table}[t!]
\centering
\caption{Performance Comparison Across Three Input Paradigms.}
\label{tab:input}
\begin{tabular}{lc}
\toprule
\textbf{Input Paradigm} & \textbf{Success in 1000 SR Processes} \\
\midrule
Indexed Data & 0 \\
General Prompt & 712 \\
Augmented Prompt & 998 \\
\bottomrule
\end{tabular}
\end{table}

\begin{figure}[t]
    \centering
    \includegraphics[width=0.5\textwidth]{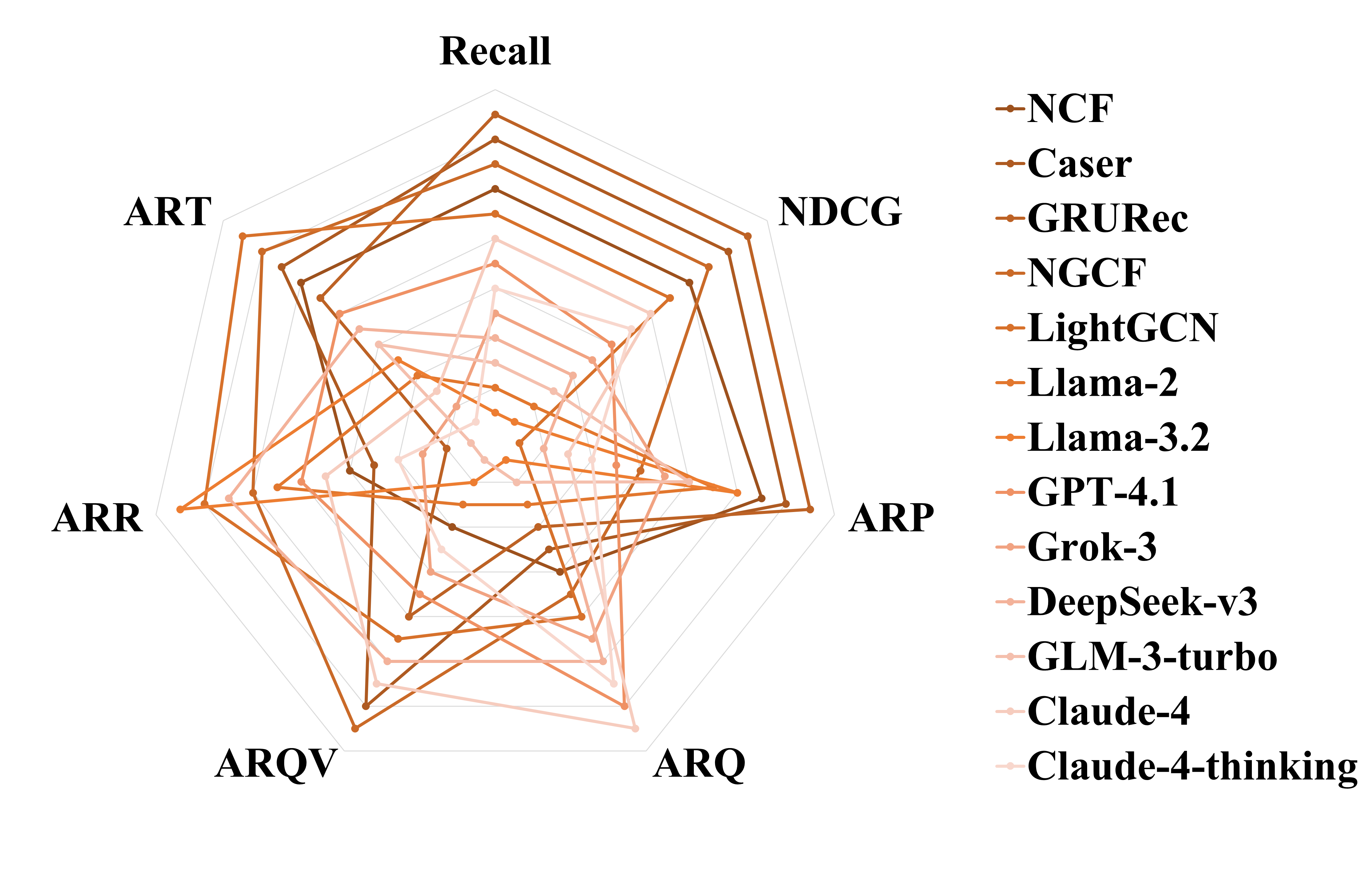}
    \caption{Overall Performance on Yelp Dataset.} 
    \label{yelp overall}
\end{figure}

\begin{figure}[t]
    \centering
    \includegraphics[width=0.5\textwidth]{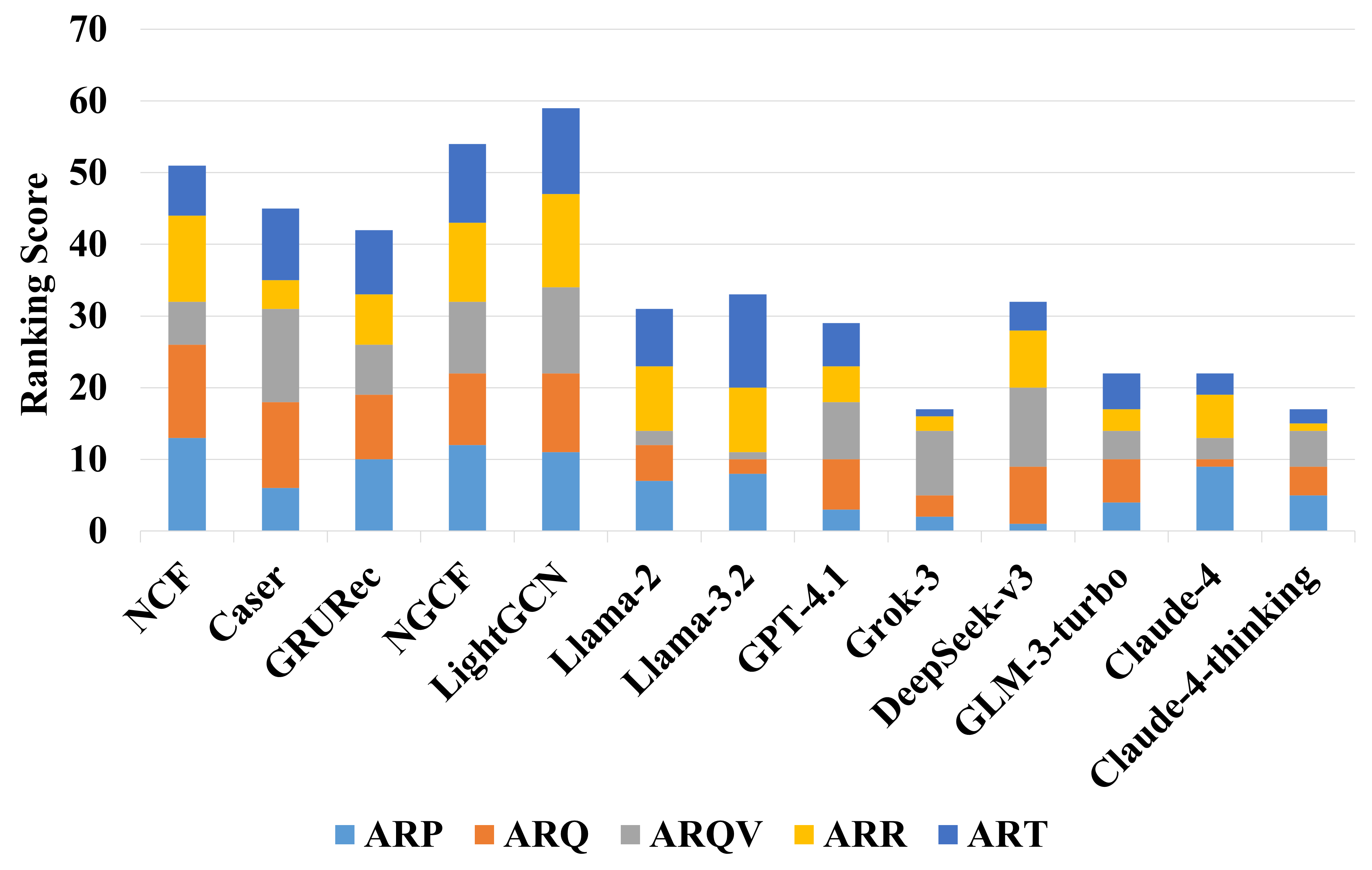}
    \caption{Performance Comparisons in Fairness, Stability and Efficiency Dimension.} 
    \label{Q2 fig}
\end{figure}

\textbf{Multi-dimensional Model Performance.} To answer \textbf{Q2}, we summarize every SR model's performance based on their ranking scores (13 to 1, descend with ranking) in each metric for normalization. As shown in Figure \ref{yelp overall}, \uline{generally, NN-SR models demonstrate better overall performance than LLM-SR models}. After summarizing SR model's multi-dimensional performance on various datasets in Table \ref{tab_dataset_advantages_symbols}, we find that \uline{LLM-SR models perform similarly with NN-SR models in accuracy dimension, while fall behind in other dimensions} as Figure \ref{Q2 fig} presents. Compared with NN-SR models, LLM-SR models' higher ARP and lower ARQ demonstrate a clear popularity-driven exposure tendency, which implies that \uline{LLM-SR models are more inclined to recommend popular items, while overlooking the item quality}. Such popularity-driven exposure tendency will cause unfair SR process to both items and users, where niche but high-quality items can't be equally exposed and users fail to receive premium SR results. The reason of this tendency may originate from LLM's reasoning capability disparity towards different information types, where LLM exhibits weaker understanding to item quality interpretation compared to item popularity. 

\textbf{Effectiveness of Unified Input Paradigm.} To answer \textbf{Q3}, we manually observe and record 1000 SR processes executed by GPT-4.1 on Yelp dataset for three rounds, checking if the model successfully executes SR process and generates requested results. The model is respectively input with original indexed data, general SR prompt with motivation, and augmented prompt after full process of prompt enhancement. Shown in Table \ref{tab:input}, the augmented prompt after prompt enhancement exhibits the best performance in facilitating successful SR process (extending approximately 40\% of general prompt's performance), which indicates that \uline{the unified input paradigm in SRBench is effective for data conversion and prompt enhancement}.

\textbf{Robustness of Prompt-extractor-coupled Extraction Mechanism.} To answer \textbf{Q4}, we manually exam 1000 pairs of output contexts and extracted results from GPT-4.1 on Yelp dataset. After the examination, only 5 pairs of output contexts and extracted results are mismatched, while the rest of output contexts are precisely extracted. \uline{Such performance demonstrates the high effectiveness and accuracy of SRBench's prompt-extractor-coupled extraction mechanism}.

\textbf{Impact of User Interaction Sequence Length.} To answer \textbf{Q5}, we additionally compare the LLM-SR model performance in few-shot (taking 5 latest interacted items orderly as user interaction sequence) and full-length mode on Yelp dataset, showing results in Figure \ref{Q5 fig}. Considering the LLM's inherent randomness, although in stability dimension, few-shot SR process exhibits higher ARQV that represents unstable recommendation quality compared to full-length SR process, in dimensions of accuracy, fairness and efficiency, few-shot SR process performances better than full-length peers, which indicates that \uline{the information contained in few-shot interaction sequence (length of 5) is enough for LLM-SR models to execute accurate, fair and efficient SR process}. Extra length in user interaction sequence may increase the data and time cost while also lower the SR performance.

\subsection{Experiment Insights}
\label{es-3}

\begin{table*}[t]
\centering
\caption{Overall Performance Comparison between Two Model Paradigms on Various Datasets.}
\label{tab_dataset_advantages_symbols}
\begin{tabular}{lcccc}
\toprule
\textbf{Dataset} & \textbf{Accuracy} & \textbf{Fairness} & \textbf{Stability} & \textbf{Efficiency} \\
ML-100K & 
NN-SR $>$ LLM-SR & NN-SR $>$ LLM-SR & NN-SR $>$ LLM-SR & NN-SR $>$ LLM-SR \\
Beauty & 
NN-SR $\approx$ LLM-SR & NN-SR $<$ LLM-SR & NN-SR $>$ LLM-SR & NN-SR $>$ LLM-SR \\
Yelp & 
NN-SR $<$ LLM-SR & NN-SR $>$ LLM-SR & NN-SR $>$ LLM-SR & NN-SR $>$ LLM-SR \\
\bottomrule
\end{tabular}

\end{table*}

\begin{itemize}

\item \textbf{Effectiveness of SRBench.} SRBench is an effective multi-dimensional benchmark for SR. Its unified input paradigm precisely converts input data with prompt enhancement, while prompt-extractor-coupled extraction mechanism reliably extracts SR results from LLM output context, aligning data between model paradigms. Combining multi-dimensional metrics with unified input paradigm and prompt-extractor-coupled extraction mechanism, SRBench enables fair and comprehensive evaluation across models and datasets. 

\begin{figure*}[t!]
    \centering
    \includegraphics[width=0.9\textwidth]{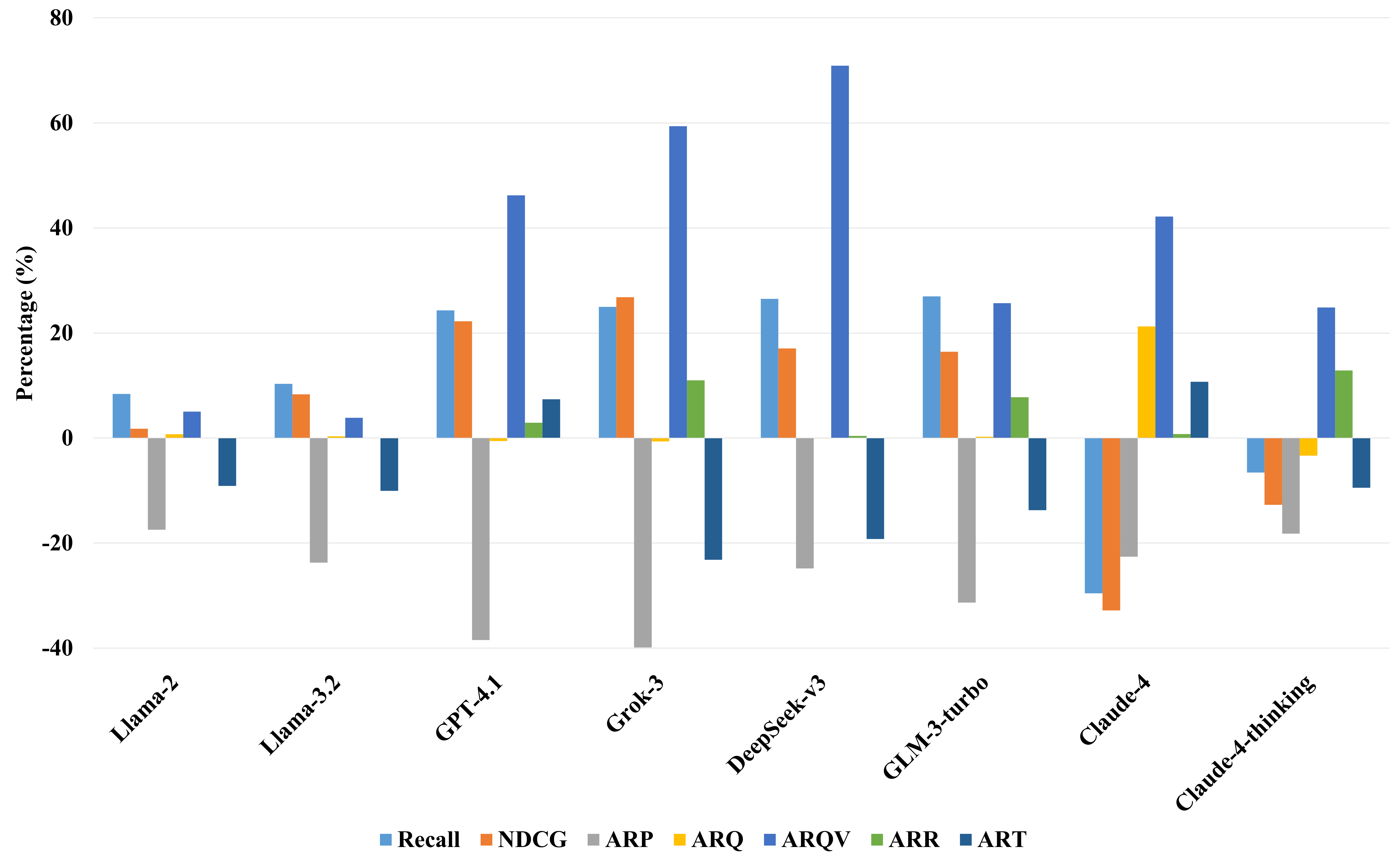}\\
    \caption{
    Performance Comparison between Few-shot and Full-length SR Process. (Percentage represents the difference that few-shot compares to full-length SR process in each metric)} 
    \label{Q5 fig}

\end{figure*} 

\item \textbf{Popularity-driven Exposure Tendency.} Compared to NN-SR models, LLM-SR models exhibits a clear popularity-driven exposure tendency, which diminishes recommendation fairness to both items and users. We believe that the tendency originates from LLM's weak understanding towards item quality information. To mitigate this issue currently, users need to carefully select and verify after receiving LLM-SR results. For future research, researchers may try apply extra prompt enhancement to remind LLM-SR models, exploring if LLM-SR models is capable of executing more quality-focused SR process.

\item \textbf{LLM-SR Stability Defect.} LLM-SR models demonstrate apparent output inconsistency compared to NN-SR models even when temperature value is set to 0, which mostly results from their inner transformer architecture. In further research, LLM-SR models require more SR training and fine-tuning process for better stability performance. 

\item \textbf{Impact of Few-shot LLM-SR Process.} Despite lower stability, the few-shot interaction sequence (length of 5) contains enough information for LLM-SR models to perform accurately, fairly and efficiently. In future LLM-SR process, user can apply the few-shot methods to lower time and throughput cost.

\item \textbf{Hallucination in LLM-SR models.} As discussed in Appendix \ref{hallu}, LLM-SR models exhibit apparent hallucination, such as Claude-Sonnet-4-thinking on Yelp dataset and Grok-3 on all datasets. To address this issue, additional validation like item ID verification is required when users are directly recommended by LLM-SR models. 

\item \textbf{Impact of LLM Thinking Mode.} Although LLM-SR model's extra thinking process improves user-perspective fairness and quality-perspective stability, it reduces the performance of LLM-SR models in other dimensions. Thus, it is still unnecessary for users to practically apply thinking mode in LLM-SR process currently.

\item \textbf{Comparison between NN-SR and LLM-SR Models.} While NN-SR models still generally outperform LLM-SR models in overall SR performance, current LLM-SR models have already achieved similar performance with NN-SR models on accuracy dimension, which reveals their huge potential in SR process. In other critical dimensions like fairness, stability and efficiency, LLM-SR models still require further improvements to generate fair, stable and efficient SR results. Inspired by NN-SR model's specially-trained neural networks, LLM-SR models require more SR-oriented trainings and fine-tunings to increase their exposure fairness, result stability and recommendation efficiency.
\end{itemize}

\section{Conclusion}

\label{conclusion}

In this paper, we propose SRBench, a comprehensive benchmark for evaluating SR models. The SRBench consists of multi-dimensional metrics, a unified input paradigm and a prompt-extractor-coupled extraction mechanism, enabling fair assessment across different model paradigms on multiple datasets. In SRBench, we evaluate 5 typical NN-SR models and 8 widely-used LLM-SR models, discovering intriguing insights like LLM-SR model's popularity-driven exposure tendency and stability defect. We hope that SRBench could be a beacon in SR model evaluation, which points the direction of comprehensive multi-dimensional evaluation framework and catalyzes more meaningful work.

\section*{Impact Statement}
\label{impact statement}
This paper presents SRBench, a benchmark for sequential recommendation that provides a holistic evaluation of LLM-SR and NN-SR models. This work aims to advance sequential recommendation research by providing a benchmark that enables rigorous, multi-dimensional evaluation of models and fair comparison across paradigms. We argue that this work is not directly linked to specific societal or ethical issues. Its contribution is mainly in highlighting the broader effects of recommendation systems across domains such as online retail, and social media.
\bibliographystyle{icml2026} 
\bibliography{ref} 
\appendix
\newpage
\onecolumn
\clearpage
\onecolumn
\label{appendix}
\appendix
\setcounter{section}{0}
\renewcommand{\thesection}{\Alph{section}}

\section*{\LARGE Supplementary Material}
\addcontentsline{toc}{section}{Supplementary Material}

\section{Additional Information}
In this section, we provide the additional information of the research process.
{\renewcommand{\arraystretch}{0.9}

\subsection{Datasets}
\label{dataset}
\begin{itemize} 
\item \textbf{MovieLens-100K} (ML-100K) \citep{harper2015movielens}, is one of the most classic benchmark dataset that contains 100,000 movie ratings from 943 users on 1,682 items, representing the entertainment domain of the SR application.

\item \textbf{Beauty} \citep{ni2019justifying} is a commonly used e-commerce dataset consisting of 371,345 user-item interactions involving 1,398 users and 32,488 products. Compared to ML-100K, it reflects typical user purchasing behaviors in marketing domain, which enriches the variety of datasets.

\item \textbf{Yelp} \citep{jendal2025yelp} includes 6,990,280 rating interactions from 851,921 users on 150,346 businesses. Compared with the other two datasets, Yelp is significantly larger in scale, which enables benchmark in large-scale SR scenarios.
\end{itemize}

\begin{table}[t]
\centering
\caption{Dataset Details}
\label{tab:dataset_details}
\begin{tabular}{lcccc}
\toprule
\textbf{Datasets} & \textbf{Ratings} & \textbf{Users} & \textbf{Items} & \textbf{Item Attributes}\\
\midrule
ML-100K & 100,000 & 943 & 1,682 & Title, Rating, Release Date, IMDB URL, Category\\
Beauty & 371,345 & 1,398 & 32,488 & Title, Overall (Rating), Brand, Rank, Description \\
Yelp & 6,990,280 & 851,921 & 150,346 & Name, Stars (Rating), Category \\
\bottomrule
\end{tabular}
\end{table}}

The preprocessing is unified to all datasets as follows: user interaction list sorted by timestamp is divided into two parts under the ratio of 9:1. The part contains 90\% of the item list is set as user historical interaction sequence, while first item of the other part is set as the true value for validation. Due to the different size of datasets, the maximum interaction sequence length is set as 50 on ML-100K and Beauty dataset, and 100 on Yelp dataset. Additionally, we excluded users with fewer than 5 interactions in the ML-100K and Beauty datasets, and fewer than 10 interactions in the Yelp dataset, ensuring the chronological order of user interaction sequence.

As summary, we evaluate SR models on three representative datasets including  ML-100K, Beauty  and Yelp in real-world settings. Table~\ref{tab:dataset_details} presents the detailed statistics for each dataset, including the number of ratings, users, items and item attributes.

\subsection{Evaluated SR Models and Settings}

\label{evaluated}
The NN-SR models are:
\begin{itemize}
\item \textbf{NCF}~\citep{he2017neural} is a classical collaborative filtering model and has been extensively used as a standard baseline in SR benchmarks. Following common evaluation mode, we include NCF to ensure comparability with previous work.

\item \textbf{Caser}~\citep{tang2018personalized} is convolution-based SR models, which explicitly model local sequential patterns through convolutional operations. We include Caser as a representative of convolutional neural network for sequence modeling.

\item \textbf{GRURec}~\citep{hidasi2018recurrent} is a widely adopted recurrent neural network-based model for recommendation tasks. It has been evaluated across multiple benchmarks as a strong baseline for capturing temporal dependencies.


\item \textbf{NGCF}~\citep{wang2019neural} and \textbf{LightGCN}~\citep{he2020lightgcn} are graph-based recommendation models that exploit user-item interaction graphs. NGCF represents early graph neural network-based methods, while LightGCN is a simplified and more efficient variant that has achieved state-of-the-art performance in SR benchmarks. These models enable a comprehensive evaluation of graph-based methods.
\end{itemize}

Based on deployment method, LLM-SR models are divided into two categories: \emph{Local LLMs} and \emph{Online LLMs}.
\begin{itemize}
\item \textbf{Local LLMs.} We consider Llama-series LLMs due to their open-source nature and widespread adoption in SR. Specifically, we locally deploy \textbf{Llama-2} and \textbf{Llama-3.2}~\citep{jayaseelan2023llama,choi2025improving}. Llama-2 serves as a widely used and well-established baseline evaluated in recommendation benchmarks, ensuring comparability with existing work. Llama-3.2 represents a more recent and stronger variant, which facilitates a comparison of different model scales within the same architecture.

\item \textbf{Online LLMs.} We select \textbf{GPT-4.1}, \textbf{Grok-3}, \textbf{DeepSeek-v3}, and \textbf{GLM-3-turbo}~\citep{zhou2025large,choi2025improving} as representative commercial LLMs, as they are widely regarded as state-of-the-art proprietary LLM-SR models in benchmarks. We also include \textbf{Claude-Sonnet-4} and its thinking mode to represent LLMs with explicit reasoning mechanisms~\citep{kusano2025revisiting}. While most selected models focus on direct generation, Claude-Sonnet-4 can test whether extra enhanced reasoning function brings additional benefits for SR, thereby complementing models without reasoning designs.
\end{itemize}
\subsection{Prompt Details}
\label{prompt details}
\begin{promptbox}{General Prompt}

A user has the following watched item history (internal Item IDs): \textbf{\{Interaction\_sequence\}}.

Based on this history, predict ONLY the top \textbf{\{Recommendation\_length\}} most likely next item IDs for this user.

\vspace{0.3cm}

\end{promptbox}

\begin{promptbox}{Augmented Prompt after Prompt Enhancement}

You are an accurate recommendation system for the \textbf{\{Dataset\_name\}}.

A user with ID \textbf{\{User\_id\}} has the following watched item history (internal Item IDs) with item information: \textbf{\{Interaction\_sequence\_with\_item\_info\}}.

Based on this history, predict ONLY the top \textbf{\{Recommendation\_length\}} most likely next item IDs for this user.

Each item ID must be an integer between 1 and \textbf{\{Total\_num\_of\_items\}}.

Output the predicted item IDs in order of likelihood, from most to least likely.

Important: Do not include any additional text, explanations, or formatting!

Output format: A single line of exactly \textbf{\{Recommendation\_length\}} comma-separated integers.

No explanation, no text, no line breaks.

Example (correct format):

42,15,301,2,104

Now generate the output:

\vspace{0.3cm}

\end{promptbox}

%

\section{Additional Experiment Results and Analysis}
In this section, we demonstrate additional experiment results and make some further analysis based on them.
\label{additional}
\subsection{Results on ML-100K Datasets}

Table~\ref{tab:100k} presents a comprehensive performance comparison of all evaluated models on the \textbf{ML-100K dataset}, a classical small-scale benchmark widely used in sequential recommendation research. Despite its small size, ML-100K remains an important tested for analyzing model behavior under relatively user-item interaction and item diversity.

\textbf{Accuracy.} On ML-100K, NN-SR models achieve stronger ranking performance compared to locally deployed LLM-SR models. GRURec attains the highest Recall@5 and NDCG@5 among NN-based methods, while graph-based models such as NGCF and LightGCN remain competitive but do not over GRURec. Online LLM-SR models, including GPT-4.1, Grok-3, and DeepSeek-v3, show varying degrees of ranking effectiveness, with GPT-4.1 achieving the best Recall@5 among LLMs, yet still falling short of the top NN-SR performance on this small-scale dataset. Llama-2 and Llama-3.2 perform the worst in terms of ranking accuracy, highlighting the impact of model scale and data diversity in smaller recommendation scenarios.

\textbf{Fairness.} NN-SR models consistently produce lower ARP values, which means less popularity bias. NCF demonstrates the lowest ARP and the highest ARQ, suggesting more balanced exposure across items while maintaining recommendation quality. Conversely, LLM-SR models exhibit higher ARP values, reflecting a tendency to over-recommend popular items, which aligns with the trends observed on larger datasets but is more pronounced here due to limited data scale.

\textbf{Stability.} The experimental results reveal differences in stability across paradigms. LightGCN shows the highest ARR among NN-SR models, indicating highly reproducible recommendation lists across repeated runs. Most LLM-SR models achieve lower ARR and higher ARQV, reflecting greater variability in both recommendation quality and output composition. Obviously, this instability stick to under deterministic generation, suggesting that the variability is intrinsic to the generation process of LLM-SR models.

\textbf{Efficiency.} NN-SR models require substantially less inference time, with LightGCN achieving the lowest ART, confirming their suitability for latency-sensitive applications. On the other hand, LLM-SR models bring higher computational costs, particularly Grok-3 and Claude-Sonnet-4-thinking, which limits their practicality for small-scale, real-time recommendation scenarios.

\begin{table*}[t]
\centering
\caption{Multi-dimensional Performance Comparison On ML-100K Dataset.}
\label{tab:100k}
\setlength{\tabcolsep}{4pt} 
\begin{tabular}{@{}c*{7}{c}@{}} 
\toprule
\multirow{2}{*}{\textbf{Model}}  & 
\multicolumn{2}{c}{\textbf{Accuracy}} & 
\multicolumn{2}{c}{\textbf{Fairness}} & 
\multicolumn{2}{c}{\textbf{Stability}} & 
\multicolumn{1}{c}{\textbf{Efficiency}} \\
\cmidrule(lr){2-3}\cmidrule(lr){4-5}\cmidrule(lr){6-7}\cmidrule(lr){8-8}
 & Recall@5 & NDCG@5 & ARP & ARQ & ARQV & ARR & ART(s) \\
\midrule
NCF & 0.0480 & 0.0303 & 22.6273 & 3.6159 & 0.1424 & 0.7415 & 0.1057 \\
Caser & 0.0637 & 0.0400 & 22.1312 & 3.5536 &  0.0770 & 0.6863 & 0.0910\\
GRURec & \textbf{0.0966} & \textbf{0.0624} & \textbf{18.6986}  & 3.5363 & 0.1098 & 0.5388 & 0.1184 \\
NGCF & 0.0630 & 0.0383 & 358.1413 & 3.7970  & \textbf{0.0745} & 0.8356 & 0.0071 \\
LightGCN & 0.0413 & 0.0265 & 469.3829 & 3.8450 & 0.1018 & 0.8813 & \textbf{0.0033} \\
\midrule
Llama-2 & 0.0037 & 0.0021 & 139.1304 & 3.4247 & 0.3273 & 0.8208 & 4.3794 \\
Llama-3.2 & 0.0035 & 0.0018 & 97.7318 & 3.2317 & 0.4672 & \textbf{0.8818} & 2.5280 \\
GPT-4.1 & 0.0293 & 0.0194 & 375.2175 & \textbf{4.0947} & 0.1245 & 0.8207 & 1.4181 \\
Grok-3 & 0.0269 & 0.0173 & 315.1565 & 3.9960 & 0.1261 & 0.6231 & 8.7538 \\
DeepSeek-v3 & 0.0194 & 0.0141 & 435.5766 & 4.0725 & 0.0907 & 0.8756 & 1.5542 \\
GLM-3-turbo & 0.0053 & 0.0034 & 162.0605 & 3.3509 & 0.5504 & 0.2830 & 2.3199 \\
Claude-Sonnet-4 & 0.0328 & 0.0219 & 416.4795 & 4.1232 & 0.0882 & 0.7518 & 6.2669 \\
Claude-Sonnet-4-thinking & 0.0286 & 0.0200 & 403.8107 & 4.0775 & 0.1331 & 0.6304 & 12.9315 \\
\midrule
Llama-2 (few-shot-5) & 0.0068 & 0.0044 & 121.2868 & 3.3787 & 0.3117 & 0.8229 & 3.2202 \\
Llama-3.2 (few-shot-5) & 0.0042 & 0.0390 & 83.3428 & 3.1817 & 0.6480 & 0.8783 & 1.5943 \\
GPT-4.1 (few-shot-5) & 0.0238 & 0.0151 & 376.6973 & 4.0791 & 0.0905 & 0.8028 & 1.2467 \\
Grok-3 (few-shot-5) & 0.0217 & 0.0130 & 316.7492 & 3.8967 & 0.1519 & 0.5670 & 9.3265 \\
DeepSeek-v3 (few-shot-5) & 0.0199 & 0.0143 & 458.2843 & 4.0928 & 0.0651 & 0.8820 & 1.4357 \\
GLM-3-turbo (few-shot-5) & 0.0083 & 0.0052 & 136.5003 & 3.3307 & 0.4511 & 0.3031 & 1.7345 \\
Claude-Sonnet-4 (few-shot-5) & 0.0276 & 0.0183 & 431.0909 & 4.1230 & 0.0575 & 0.7738 & 7.8404 \\
Claude-Sonnet-4-thinking (few-shot-5) & 0.0246 & 0.0168 & 430.1300 & 4.0944 & 0.1008 & 0.6650 & 12.1489 \\
\bottomrule 
\end{tabular}
\end{table*}

These results reflect a stable balance of advantages and limitations across paradigms on ML-100K. Specifically, NN-SR models provide more balanced performance across accuracy, fairness, stability, and efficiency, but LLM-SR models demonstrate competitive in certain accuracy metrics, generally fall in fairness, stability, and efficiency on small datasets. This highlights the importance of multi-dimensional evaluation on classical benchmarks to obtain a complete understanding of model behavior.

\subsection{Results on Beauty Datasets}

\begin{table*}[t]
\centering
\caption{Multi-dimensional Performance Comparison On Beauty Dataset.}
\label{tab:beauty}
\setlength{\tabcolsep}{4pt} 
\begin{tabular}{c*{7}{c}} 
\toprule
\multirow{2}{*}{\textbf{Model}} & 
\multicolumn{2}{c}{\textbf{Accuracy}} & 
\multicolumn{2}{c}{\textbf{Fairness}} &  
\multicolumn{2}{c}{\textbf{Stability}} & 
\multicolumn{1}{c}{\textbf{Efficiency}} \\
\cmidrule(lr){2-3}\cmidrule(lr){4-5}\cmidrule(lr){6-7}\cmidrule(lr){8-8}
 & Recall@5 & NDCG@5 & ARP & ARQ & ARQV & ARR & ART(s) \\
\midrule
NCF & 0.4549  & 0.3441 &128.8000  & 4.6570 & 0.1065 &\textbf{0.9000}  & 0.0754 \\
Caser &  0.4964 & 0.4755 & 382.0719 &  \textbf{4.6842} & \textbf{0.0693} & 0.3661 & 0.4759\\
GRURec & \textbf{0.5207}&  \textbf{0.4992} & 589.5371 & 4.4383 & 0.2058  & 0.6368  & 0.6575\\
NGCF &  0.1994 & 0.0923  & 544.6045 & 4.4731 & 0.1820& 0.7215  & 0.0166\\
LightGCN & 0.2504 & 0.1391 &807.2668 & 4.4448& 0.1705 & 0.7818 & \textbf{0.0042}\\
\midrule
Llama-2 &  0.0665 & 0.0381 & 122.9745 & 4.4149 &0.7885  & 0.4499 & 0.5971 \\
Llama-3.2 & 0.2031 & 0.1072 & 451.2622 &4.6133 & 0.3896 & 0.4504 & 0.2306 \\
GPT-4.1 & 0.3415 & 0.2301 & 283.1251 & 4.4850 & 0.4188 & 0.4268 & 2.0263 \\
Grok-3 & 0.1800 & 0.1454 & 210.1240 & 4.4351 & 0.4667 & 0.3583 & 5.5472 \\
DeepSeek-v3 & 0.2150 & 0.1873 & 207.6472 & 4.4109 & 0.4883 & 0.4427 & 1.4078 \\
GLM-3-turbo & 0.1248 & 0.0894 & 139.8540 & 4.0288 & 0.6186 & 0.2536 & 2.3016 \\
Claude-Sonnet-4 & 0.3265 & 0.2314 & \textbf{50.0571} & 4.3510 & 0.6153 & 0.3967 & 7.7099 \\
Claude-Sonnet-4-thinking & 0.2462 & 0.1831 & 93.6647 & 4.4348 & 0.5034 & 0.2604 & 10.1674 \\
\midrule
Llama-2 (few-shot-5) & 0.0680 & 0.0385 & 123.1160 & 4.4638 & 0.6459 & 0.4500 & 0.4868 \\
Llama-3.2 (few-shot-5) & 0.2103 & 0.1117 & 453.7549 & 4.6127 & 0.3917 & 0.4504 & 0.2111 \\
GPT-4.1 (few-shot-5) & 0.3364 & 0.2296 & 277.7846 & 4.4767 & 0.4208 & 0.4247 & 1.1043 \\
Grok-3 (few-shot-5) & 0.2023 & 0.1541 & 182.9654 & 4.3968 & 0.5259 & 0.3009 & 8.7304 \\
DeepSeek-v3 (few-shot-5) & 0.2209 & 0.1901 & 208.1483 & 4.4056 & 0.4931 & 0.4464 & 1.3504 \\
GLM-3-turbo (few-shot-5) & 0.1296 & 0.0913 & 116.3415 & 3.7462 & 0.7152 & 0.2572 & 2.3601 \\
Claude-Sonnet-4 (few-shot-5) & 0.3810 & 0.2568 & 44.8301 & 4.4098 & 0.5704 & 0.4407 & 2.5265 \\
Claude-Sonnet-4-thinking (few-shot-5) & 0.1792 & 0.1517 & 63.5498 & 4.3754 & 0.4801 & 0.2288 & 9.4569 \\
\bottomrule 
\end{tabular}
\end{table*}  

Table~\ref{tab:beauty} presents the evaluation results on the \textbf{Beauty dataset}, which captures user interactions in e-commerce domain with medium scale compared to ML-100K and Yelp, allowing us to examine model behavior under typical commercial recommendation scenarios.  

\textbf{Accuracy.} NN-SR models generally achieve best ranking results, with GRURec attaining the highest Recall@5 and NDCG@5 among neural networks baselines. Graph-based methods such as NGCF and LightGCN remain competitive, yet their ranking performance doesn't surpass GRURec. It's interesting that some online LLM-SR models including GPT-4.1 and DeepSeek-v3 show noticeable accuracy improvements over locally deployed LLMs like Llama-2 and Llama-3.2, although they still fail to consistently outperform the high performance NN-SR models. This demonstrates that while LLM-SR models can generate high-quality recommendations, their advantages are less evident on datasets of moderate scale.  

\textbf{Fairness.} Considering exposure equity, NN-SR models maintain relatively lower ARP values, which implies reduced popularity bias. NCF achieves the lowest ARP and highest ARQ, suggesting a fairer distribution of recommendations across items. On the other hand, higher ARP values are observed for LLM-SR models, which reveals a tendency to favor popular items and a trade-off between ranking accuracy and fairness. In a words, the models that perform best in ranking do not automatically ensure balanced item exposure.  

\textbf{Stability.} Stability analysis reveals that NN-SR models like LightGCN, produce highly consistent recommendation lists across repeated runs, highlighting predictable and reliable behavior. Conversely, LLM-SR models exhibit greater variability, with noticeable instability in both recommendation composition and ranking quality. This instability persists even under deterministic generation, suggesting that the variability arises from the LLM generation mechanism itself rather than random factors. Although LLM-SR models can occasionally outperform NN-SR models in accuracy, their reproducibility remains less certain.  

\textbf{Efficiency.} In terms of inference time, NN-SR models are markedly faster with LightGCN achieving the lowest ART that demonstrates suitability for time-sensitive deployment. In contrast, LLM-SR models shows higher computational costs, particularly Grok-3 and Claude-Sonnet-4-thinking, which constrains their practical use in real-time recommendation settings. While achieving higher accuracy, LLM-SR models may exhibit disproportionately high computational costs.

In summary, these results reveal the balance of strengths and weaknesses across model paradigms on the Beauty dataset. While NN-SR models provide stable, fair, and efficient performance across dimensions, LLM-SR models, despite occasionally achieving competitive ranking accuracy, generally lag in fairness, stability, and efficiency under this medium-scale scenario. This highlights the necessity of multi-dimensional evaluation even in datasets that are not as large as Yelp to obtain a comprehensive understanding of model behavior.

\subsection{Hallucination of LLM-SR Models}
\label{hallu}
\begin{table*}[t]
\centering
\caption{Hallucination Rates of Different LLMs on Various Datasets.}
\label{tab:hallucination_rates}
\setlength{\tabcolsep}{5pt}
\begin{tabular}{l *{6}{c}}
\toprule
\multirow{2}{*}{\textbf{Model}} & \multicolumn{2}{c}{\textbf{ML-100K}} & \multicolumn{2}{c}{\textbf{Beauty}} & \multicolumn{2}{c}{\textbf{Yelp}} \\
\cmidrule(lr){2-3} \cmidrule(lr){4-5} \cmidrule(lr){6-7}
 & \textbf{full} & \textbf{few-shot-5} & \textbf{full} & \textbf{few-shot-5} & \textbf{full} & \textbf{few-shot-5} \\
\midrule
Llama-2 & 0.0002 &  0.0007 & 0.0093 & 0.0003  & 0.0002 & 0.0003 \\
Llama-3.2 & 0.0002 &  0.0007 & 0.0003 &0.0744 & 0.0009  & 0.0022 \\
GPT-4.1 & 0.0002 & 0.0000 & 0.0007 &0.0010 & 0.0015 & 0.0048 \\
Grok-3 & 0.0214 &  0.0196 & 0.0057 & 0.0075 & 0.0236 & 0.0135 \\
DeepSeek-v3 & 0.0000 & 0.0000 & 0.0025 & 0.0026 & 0.0011 & 0.0103 \\  
GLM-3-turbo & 0.0001 & 0.0003 & 0.1088 & 0.1658 & 0.0081 & 0.0069 \\
Claude-Sonnet-4 & 0.0003 & 0.0008 & 0.0103 & 0.0014 & 0.1752  & 0.0006 \\
Claude-Sonnet-4-thinking & 0.0010 & 0.0001 & 0.0088 & 0.0269 & 0.0087 & 0.0360 \\
\bottomrule
\end{tabular}
\end{table*}

To evaluate the effectiveness of LLM-generated recommendations, we measure the hallucination rates of multiple LLMs across three datasets ML-100K, Beauty, and Yelp, under both full-length and few-shot (5-shot) interaction settings. Hallucinations are defined as items predicted by the model that don't exist in the ground-truth item set.

\begin{figure*}[t]     
    \centering
    \begin{subfigure}[c]{0.40\textwidth}
        \centering
        \includegraphics[width=\textwidth]{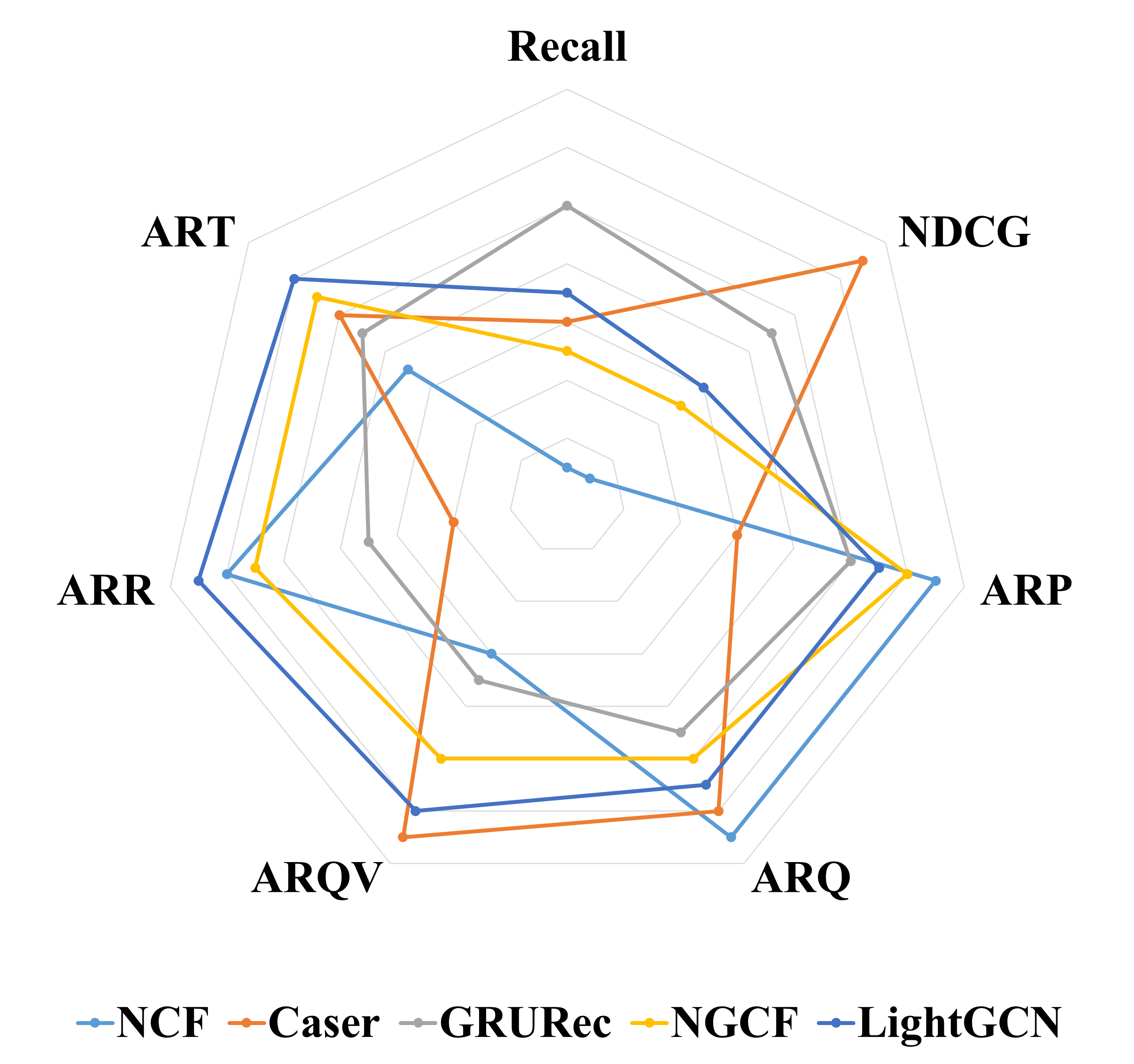}
        \label{yelp_nn}
    \end{subfigure}
    \begin{subfigure}[c]{0.55\textwidth}
        \centering
        \includegraphics[width=\textwidth]{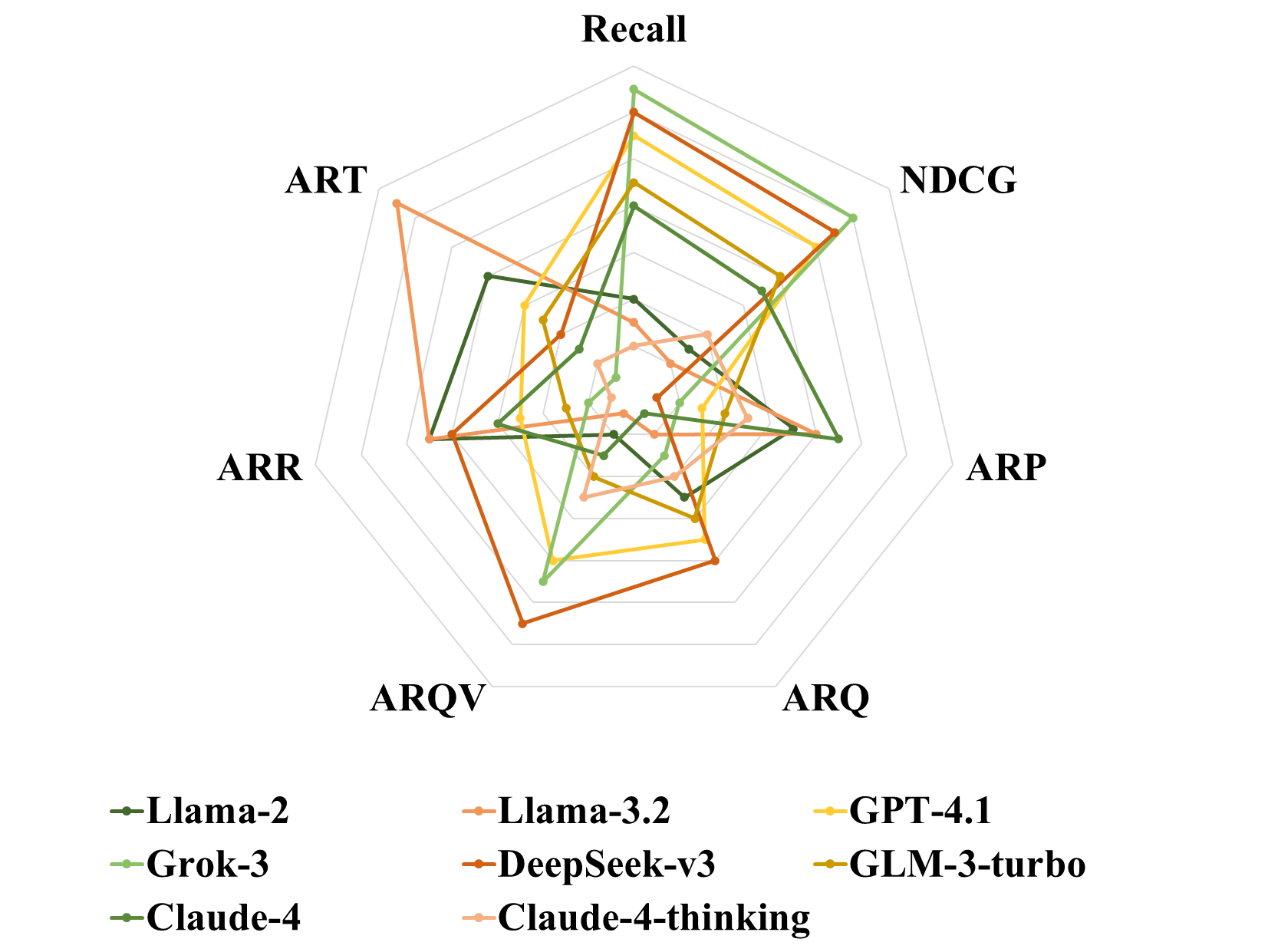}
        \label{yelp_llm}
    \end{subfigure}
        \hfill 
    \caption{Overall Performance Comparison on Yelp Dataset: 
     (a) NN-SR models and (b) LLM-SR models.}
    \label{fig_yelp_overall} 
\end{figure*}  

Table~\ref{tab:hallucination_rates} summarizes the observed hallucination rates. Specifically, locally deployed models such as Llama-2 and Llama-3.2 exhibit minimal hallucinations across all datasets in full-length settings, with rates generally below 0.01\%. However, in the few-shot configuration, Llama-3.2 shows elevated hallucinations on the Beauty dataset (7.44\%), which means sensitivity to limited user context. Online LLMs including GPT-4.1 and DeepSeek-v3 maintain consistently low hallucination rates, indicating robust generalization across datasets. In contrast, Models such as GLM-3-turbo and Claude-Sonnet-4 exhibit substantially higher hallucination rates under few-shot settings, reaching up to 16.58\% on the Beauty dataset and 17.52\% on Yelp.

These results highlight a clear divergence between open-source LLMs and commercial or larger-scale models in terms of output reliability. While certain online LLMs achieve both high recommendation accuracy and low hallucination rates, others, especially those incorporating complex reasoning or scaling mechanisms, are prone to generating invalid items when exposed to limited interaction histories. This underscores the importance of content validation mechanisms in LLM-based sequential recommendation systems to ensure the correctness and reliability of predictions.

\subsection{Detailed Comparison of Performance between Model Paradigms}

\begin{figure*}[t]     
    \centering
    \begin{subfigure}[c]{0.40\textwidth}
        \centering
        \includegraphics[width=\textwidth]{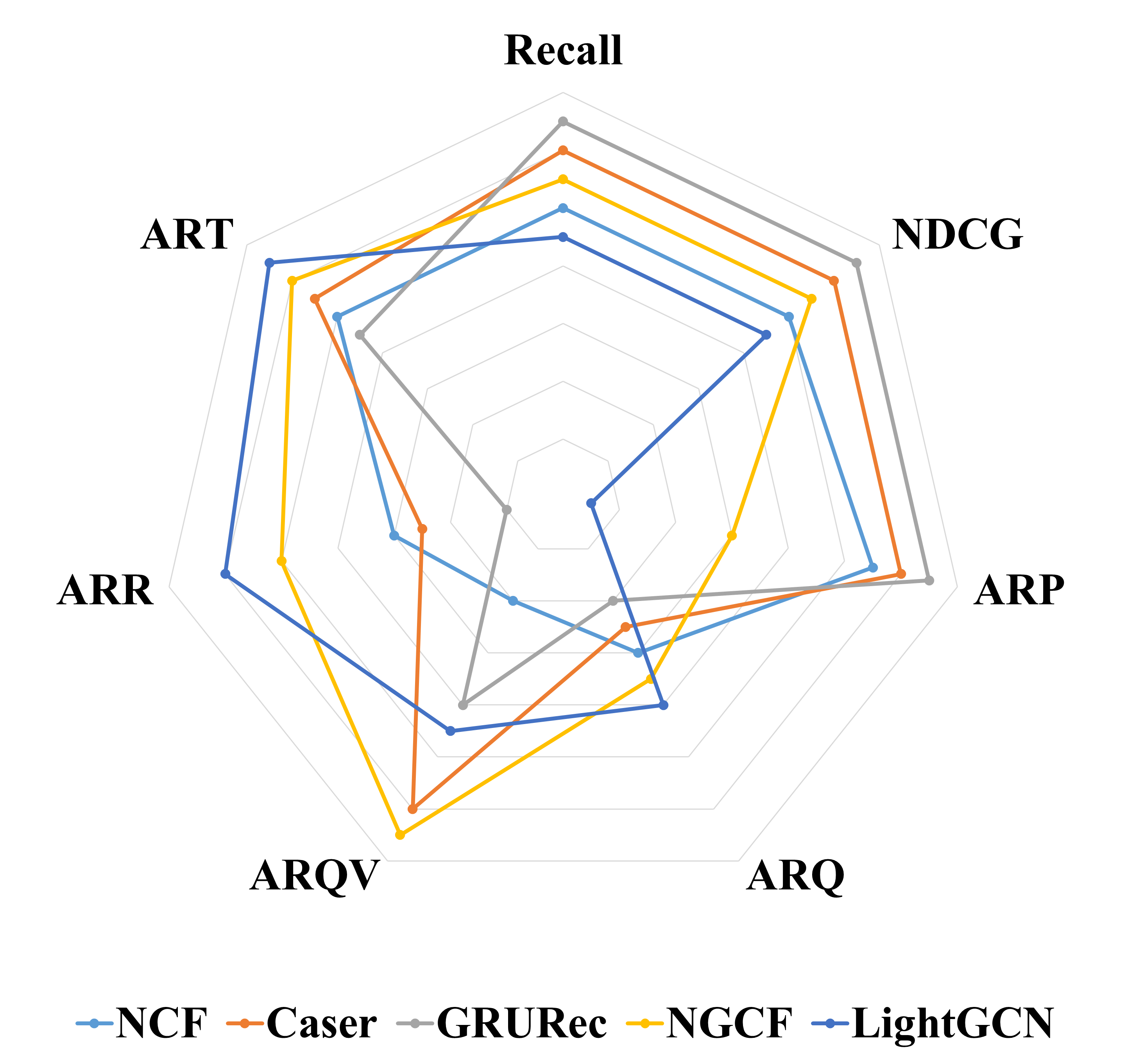}
        \label{100k_nn}
    \end{subfigure}
    \begin{subfigure}[c]{0.55\textwidth}
        \centering
        \includegraphics[width=\textwidth]{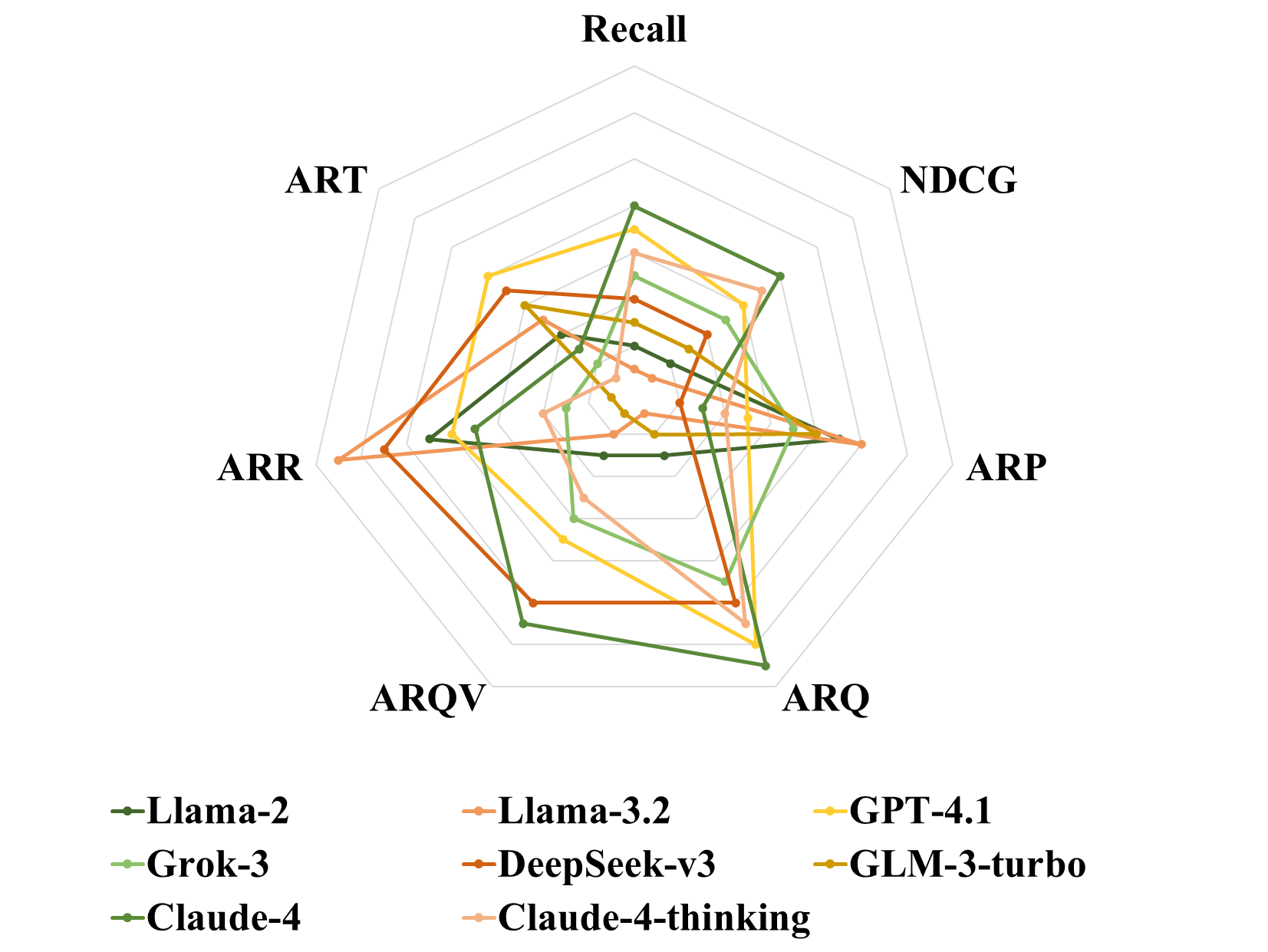}
        \label{100k_llm}
    \end{subfigure}
        \hfill 
    \caption{Overall Performance Comparison on ML-100K Dataset: 
     (a) NN-SR models and (b) LLM-SR models.}
    \label{fig_100k_overall} 
\end{figure*}  

\begin{figure*}[t]     
    \centering
    \begin{subfigure}[c]{0.40\textwidth}
        \centering
        \includegraphics[width=\textwidth]{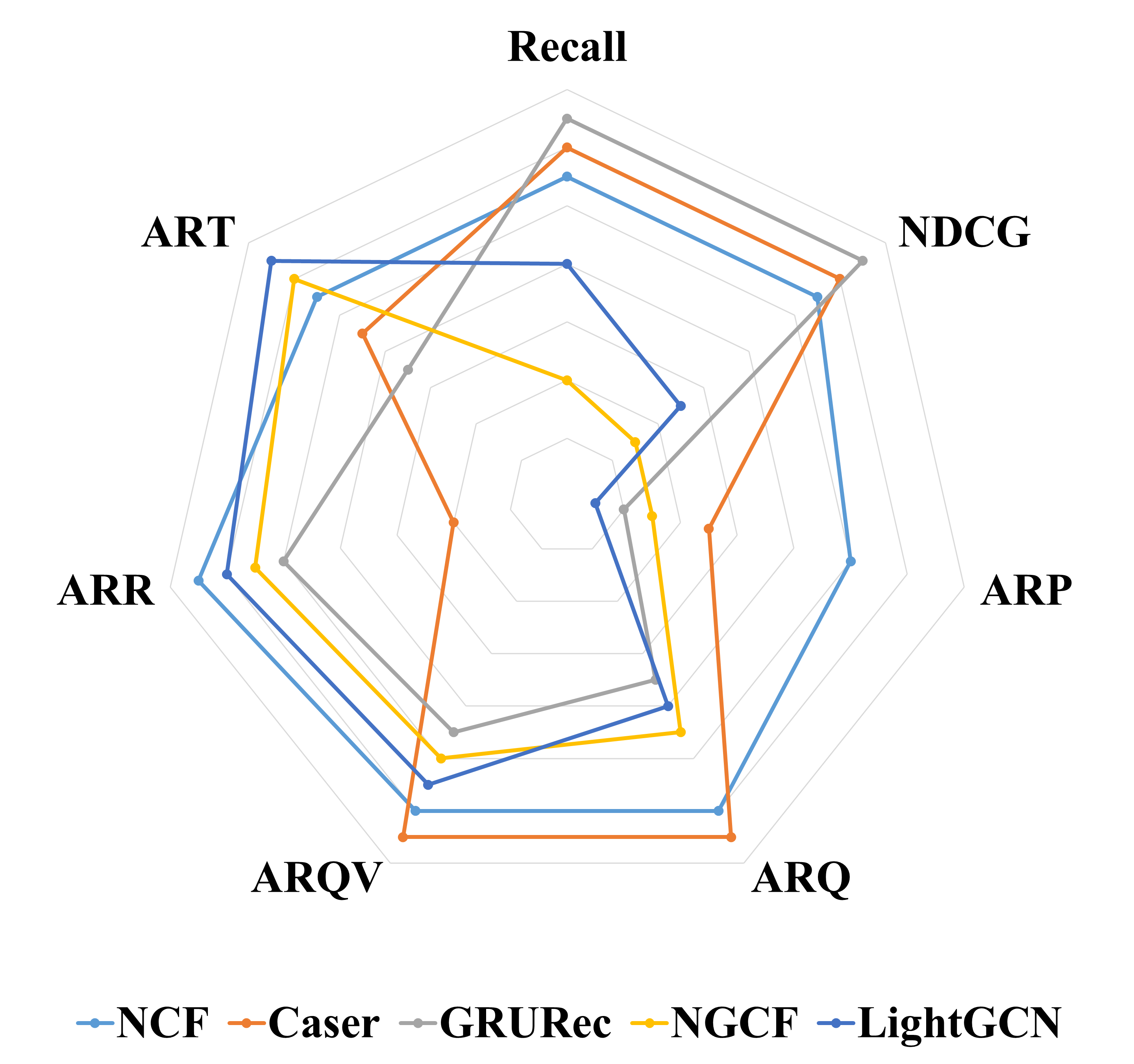}
        \label{beauty_nn}
    \end{subfigure}
    \begin{subfigure}[c]{0.55\textwidth}
        \centering
        \includegraphics[width=\textwidth]{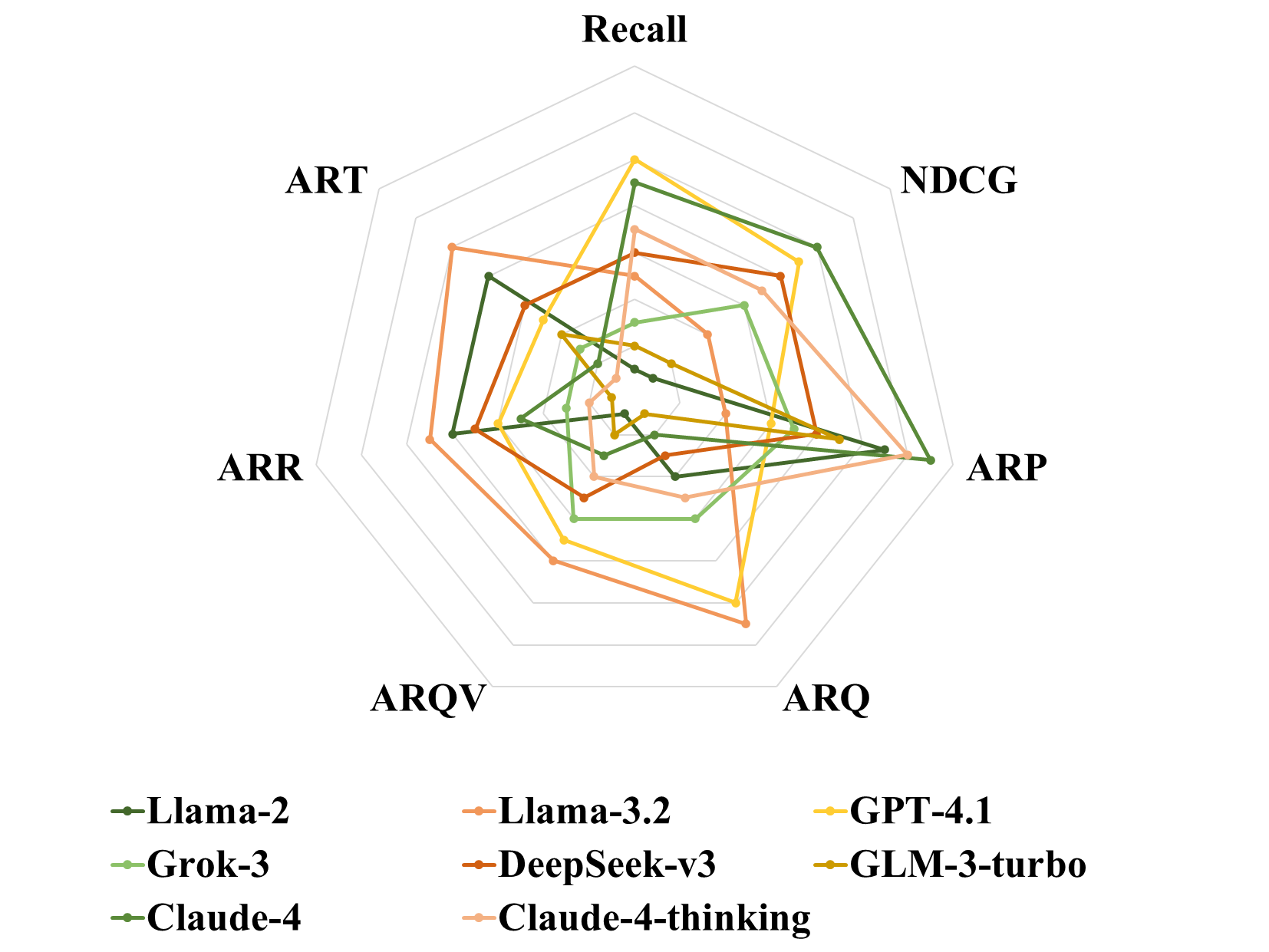}
        \label{beauty_llm}
    \end{subfigure}
        \hfill 
    \caption{Overall Performance Comparison on Beauty Dataset: 
     (a) NN-SR models and (b) LLM-SR models.}
    \label{fig_beauty_overall} 
\end{figure*} 

Figure ~\ref{fig_yelp_overall} presents a multi-dimensional comparison between NN-SR models and LLM-SR models on the Yelp dataset. An obvious divergence in performance can be observed across paradigms. NN-SR models demonstrate more balanced behavior across accuracy, fairness, stability, and efficiency, with graph-based approaches such as LightGCN achieving strong stability and low inference latency. In contrast, LLM-SR models exhibit higher variability across metrics, with certain models achieving competitive or even superior performance in accuracy or fairness dimensions, yet often at the cost of reduced stability and substantially higher computational cost.

In particular, LLM-SR models show a tendency toward popularity-driven exposure, as shown by increased ARP values, while their ARQV and ARR scores express less consistent recommendation outputs across repeated runs. Moreover, inference time (ART) remains significantly higher for LLM-SR approaches, highlighting a practical deployment challenge in latency-sensitive recommendation scenarios. These findings suggest that while LLM-SR models possess distinct advantages in semantic understanding and context reasoning, NN-SR models currently provide more reliable and efficient solutions for large-scale real-world recommendation systems.

Figure~\ref{fig_100k_overall} provides a multi-dimensional evaluation contrasting NN-SR and LLM-SR paradigms on the ML-100K dataset. LLM-SR models reveal irregular behavior across evaluation dimensions, but NN-SR approaches maintain comparatively stable performance in accuracy, fairness, stability, and efficiency. In particular, NN-SR models such as GRURec are found to achieve better ranking effectiveness, and graph-based methods like LightGCN are characterized by consistent outputs and minimal inference cost. By contrast, although LLM-SR models demonstrate significant capability in capturing ranking relevance, these gains are frequently accompanied by compromised robustness and increased computational resource.

A notable tendency toward popularity-oriented exposure is also observed in LLM-SR results, as evidenced by high ARP values, whereas NN-SR models recommendation are more balanced across items. Meanwhile, higher ARQV and lower ARR scores indicate that LLM-SR outputs are less reproducible across repeated runs, which means recommendation stability is not inherently guaranteed under LLM-SR paradigms. In addition, substantially longer inference times (ART) are suffered by LLM-SR models, which makes their deployment less practical in latency-constrained environments. Therefore, these observations mean that LLM-SR models benefit from richer semantic knowledge and contextual reasoning capacity, while NN-SR models presently offer a more dependable and computationally efficient solution for small-scale recommendation scenarios.

Figure~\ref{fig_beauty_overall} presents a multi-dimensional evaluation of NN-SR and LLM-SR models on the Beauty dataset. Following observations on on ML-100K, NN-SR models exhibit stable behavior across accuracy, fairness, stability, and efficiency metrics. Specifically, GRURec achieves better ranking effectiveness, and graph-based methods such as LightGCN maintain consistent outputs with minimal inference cost. On the contrary, LLM-SR models show notable irregularities across evaluation dimensions, with high ARP values demonstrate a tendency toward popularity-driven recommendations, while ARQV and ARR scores suggest lower reproducibility across repeated runs. Additionally, ART for LLM-SR models remain substantially higher, highlighting deployment challenges in latency-sensitive scenarios. Overall, the results suggest that LLM-SR models benefit from enhanced semantic understanding and context reasoning, whereas NN-SR approaches deliver more consistent and computationally efficient performance on the Beauty dataset.

With the three datasets ML-100K, Beauty, and Yelp are evaluated, the proposed SRBench framework reveals consistent trends and dataset-specific differences. ML-100K, a relatively small-scale movie rating dataset, exhibits stable NN-SR performance across accuracy, fairness, stability, and efficiency metrics, while LLM-SR models demonstrate irregular behavior and high computational cost. The Beauty dataset, with common size and emphasizing consumer products, shows similar patterns with ML-100K: NN-SR models such as GRURec and LightGCN maintain reliable ranking effectiveness and low inference latency, whereas LLM-SR approaches suffer from popularity bias and lower reproducibility. For the large-scale Yelp dataset, NN-SR models provide reliable and efficient performance, but LLM-SR models show greater variability across fairness, stability, and efficiency metrics despite leveraging semantic reasoning. Collectively, these findings suggest that while LLM-SR models leverage richer contextual understanding, NN-SR models currently offer more consistent and practical performance. 



\clearpage


\end{document}